\documentclass{article}

\usepackage{PRIMEarxiv}

\usepackage[utf8]{inputenc} % allow utf-8 input
\usepackage[T1]{fontenc}    % use 8-bit T1 fonts
\usepackage{hyperref}       % hyperlinks
\usepackage{url}            % simple URL typesetting
\usepackage{booktabs}       % professional-quality tables
\usepackage{amsfonts}       % blackboard math symbols
\usepackage{nicefrac}       % compact symbols for 1/2, etc.
\usepackage{microtype}      % microtypography
\usepackage{lipsum}
\usepackage{fancyhdr}       % header
\usepackage{graphicx}       % graphics
\graphicspath{{media/}}     % organize your images and other figures under media/ folder
\usepackage{graphicx}
\usepackage{subcaption}
\usepackage{multirow}
\usepackage{enumitem}
\usepackage{caption}
\usepackage{float}
\usepackage{amsmath}

\title{Enabling Versatile Human Sensing through Wi-Fi Beamforming Feedback
}

\author{
 Jingzhe Zhang \\
  University of South Florida\\
  \texttt{jingzhe@usf.edu} \\
   \And
   Peiliang Wang \\
  University of South Florida\\
  \texttt{peiliangwang@usf.edu} \\
  \And
 Yitong Shen \\
  University of South Florida\\
  \texttt{shen202@usf.edu} \\
  \And
 Ning Wang \\
  University of South Florida\\
  \texttt{ningw@usf.edu} \\
\And
 Mingzhe Chen \\
  University of Miami\\
  \texttt{mingzhe.chen@miami.edu} \\
\And
 Yili Ren \\
  University of South Florida\\
  \texttt{yiliren@usf.edu} \\
}

\begin{document}
\maketitle

\begin{abstract}
Wi-Fi sensing enables contactless human sensing, while most existing systems rely on Channel State Information, which is accessible only on a limited range of Wi-Fi chipsets.
In contrast, Beamforming Feedback Information (BFI) is standardized in modern Wi-Fi and readily accessible on a wide range of commodity devices, offering new opportunities for human sensing.
In this paper, we present BeamLens, a Wi-Fi sensing system that extracts fine-grained spatial information from BFI to enable versatile human sensing.
Specifically, we jointly estimate the angle of departure and relative time of flight from BFI to extract fine-grained spatial information and leverage two AP-STA links to capture complementary azimuth and elevation views.
We further design a sparse physical antenna array and construct a uniform virtual antenna array to enhance spatial resolution.
Moreover, we fuse complementary views through self-supervised learning, combining temporally aware cross-view contrastive learning and temporal masked reconstruction to learn a versatile backbone for diverse human sensing tasks.
Extensive evaluations demonstrate that BeamLens achieves 90.16\% accuracy in human activity recognition, a mean per-joint position error of 8.64\,cm in 3D human pose estimation, and 70.64\% intersection over union in human silhouette estimation, while exhibiting strong generalization to unseen environments and subjects.
\end{abstract}

% keywords can be removed

\section{Introduction}
% P1: WiFi sensing background & value
% P2: CSI-based systems need special hardware
% P3: introduce BFI, standard & passively capturable
% P4: present BeamLens, challenge 1
% P5: challenge 2 - few antennas, low resolution - sparse array
% P6: challenge 3 - single-task/env, label-hungry - SSL backbone
% P7: experiment summary
% P8: contributions

%Wi-Fi sensing has drawn growing interest in recent years as a contactless approach to human sensing using wireless signals already present in everyday environments. 

Wi-Fi sensing has attracted increasing attention in recent years as a contactless approach to human sensing that reuses existing Wi-Fi infrastructure.
As Wi-Fi signals propagate through the environment, they interact with the human body through reflection and scattering, producing signal variations that capture human motion and enable Wi-Fi-based human sensing~\cite{tan2022commodity, ma2019wifi}. This capability has enabled a wide range of applications, including human activity recognition~\cite{wang2014eyes, wang2015understanding, jiang2018towards}, human pose estimation~\cite{ren2021winect, ren2022gopose}, vital-sign monitoring~\cite{liu2015tracking, zeng2019farsense}, and gesture recognition~\cite{zhang2021widar3, abdelnasser2015wigest, li2016wifinger}, extending the role of Wi-Fi beyond wireless communication. With Wi-Fi devices already ubiquitous in homes, offices, and public spaces and their numbers continuing to grow with the expansion of the Internet of Things (IoT), Wi-Fi sensing holds significant potential for large-scale deployment.

Existing Wi-Fi sensing systems mainly rely on Channel State Information (CSI). However, the limited accessibility of CSI on commodity Wi-Fi devices leaves a significant gap toward achieving truly ubiquitous Wi-Fi sensing. This is because CSI extraction is supported by only a limited set of Wi-Fi chipsets and network interface cards (NICs)~\cite{wu2023enabling}, such as the Intel 5300~\cite{halperin2011tool} and Atheros AR9580~\cite{xie2015precise}, many of which belong to older Wi-Fi generations, such as Wi-Fi 4 (e.g., IEEE 802.11n). Moreover, CSI extraction requires specialized drivers, firmware modifications, or CSI extraction tools~\cite{halperin2011tool,jiang2021eliminating,gringoli2019free}, further limiting its accessibility on commodity Wi-Fi devices. Consequently, fewer than 6\% of current Wi-Fi devices support CSI extraction~\cite{wu2023enabling}, significantly limiting the scalability and widespread deployment of Wi-Fi sensing.

Recently, Beamforming Feedback Information (BFI), originally designed to enhance Wi-Fi communication performance, has emerged as a more accessible source of wireless channel information, opening new opportunities for ubiquitous Wi-Fi sensing~\cite{wu2023enabling, yi2024bfmsense}.
This is driven by two key advantages.
First, BFI is standardized in modern Wi-Fi generations, including Wi-Fi 5, 6, and 7 (i.e., IEEE 802.11ac/ax/be~\cite{ieee80211ac2013, ieee80211ax2021, 11090080}), making it widely supported by commodity Wi-Fi devices.
Second, BFI is transmitted in unencrypted feedback frames, allowing it to be readily captured using commodity Wi-Fi devices (e.g., a laptop) and widely used packet analysis tools (e.g., Wireshark), without requiring firmware or driver modifications to the devices~\cite{haque2023wi, xiao2025lend}.
%Third, although BFI contains only a compressed and partial representation of wireless channel information, it still preserves spatial characteristics of signal propagation that reflect changes caused by human movements~\cite{itahara2022beamforming, yi2024bfmsense, li2026bfmscan}.
However, research on BFI-based human sensing remains limited. Some existing systems rely on coarse-grained features (e.g., temporal variations~\cite{kondo2022bi, yi2024bfmsense} and statistical characteristics of BFI measurements~\cite{chen2024ship, kato2022cbr}), primarily supporting basic sensing tasks such as motion detection~\cite{suraweera2019stand, chen2024ship} and activity recognition~\cite{haque2025beamsense, haque2025si}. Moreover, many existing systems are limited to task-specific signal processing and feature extraction methods tailored to individual sensing tasks~\cite{xiao2025lend, hu2024m, haque2024bfa, haque2025si}, hindering their adaptability to diverse sensing applications. Consequently, the potential of extracting fine-grained information from BFI to enable versatile human sensing remains largely unexplored.

In this work, we present BeamLens, a novel Wi-Fi sensing system that exploits fine-grained spatial information embedded in BFI to enable versatile human sensing, including 3D human pose estimation, human silhouette estimation, and human activity recognition.
%system that leverages BFI to extract fine-grained spatial information for multi-task Wi-Fi sensing. 
However, realizing such a system is nontrivial and poses several challenges. First, BFI provides only a compressed and quantized representation of the wireless channel~\cite{ieee80211ac2013, ieee80211ax2021, wu2023enabling, yi2024bfmsense}, inherently limiting the human-related information it retains. This raises a fundamental question: To what extent can fine-grained spatial information about the human body be extracted from BFI?
To answer this question, we carefully analyze what channel information is preserved in BFI. BFI still retains transmit-side spatial information and frequency-dependent information across multiple subcarriers~\cite{itahara2022beamforming, kondo2022bi}. Building on these observations, we leverage the spatial diversity across multiple transmitting antennas and the frequency diversity across subcarriers to jointly estimate the angle of departure (AoD) and relative time of flight (RToF), thereby extracting fine-grained angle-distance information for human sensing.
%fine-grained human sensing requires fine-grained spatial information to characterize human motion. BFI provides only a compressed and partial representation of the wireless channel, primarily retaining transmit-side spatial information. Although BFI can be used to estimate the angle of departure (AoD), angular information alone provides only coarse-grained spatial information, making it difficult to distinguish signal paths with similar departure angles. To address this challenge, BeamLens exploits both the spatial diversity across multiple transmitting antennas and the frequency diversity across numerous subcarriers to jointly estimate AoD and relative time of flight (RToF), thereby enabling finer-grained spatial characterization by distinguishing signal paths with similar departure angles but different RToF values. 
Furthermore, to better capture the human body from complementary perspectives, BeamLens employs two orthogonally oriented access point (AP)-station (STA) links to simultaneously capture azimuth and elevation views, corresponding to the horizontal and vertical angular dimensions, respectively. By combining AoD-RToF information from both views, BeamLens captures comprehensive spatial information about the human body.

Second, the spatial resolution of BFI-based Wi-Fi sensing is inherently constrained by the limited number of antennas and the small antenna array aperture of commodity Wi-Fi devices. 
For example, commodity Wi-Fi devices (e.g., NETGEAR R7000 and R7800) typically employ only three or four antennas.
%, resulting in a limited array aperture and consequently low spatial resolution. 
To address this challenge, we rearrange the antenna array layout based on the minimum redundancy principle~\cite{moffet1968minimum}, forming a sparse physical antenna array with an enlarged antenna array aperture without requiring additional antennas or hardware.
However, despite the increased aperture, the number of physical antennas remains unchanged, limiting the number of propagation paths that can be resolved~\cite{schmidt1986multiple, kotaru2015spotfi}. To overcome this limitation, we further construct a uniform virtual antenna array from the sparse physical antenna array, increasing the effective degrees of freedom and enabling the separation of more multipath components.
Together, the sparse physical antenna array and the uniform virtual antenna array enable BeamLens to further enhance the spatial resolution.

Third, with complementary spatial information captured from two views and temporal information obtained through continuous BFI measurements, another challenge is how to effectively fuse them into a unified representation that supports diverse sensing tasks. To address this challenge, BeamLens employs a self-supervised learning framework that jointly exploits cross-view correspondence and temporal consistency. Specifically, BFI measurements captured simultaneously from the two views naturally form positive pairs, allowing cross-view contrastive learning to bring their representations closer while separating negative pairs. In addition, temporally neighboring BFI measurements are treated as weighted soft positive pairs to capture the continuity of human movements. BeamLens further incorporates temporal masked reconstruction to learn temporal dependencies by reconstructing missing observations from their surrounding context. These learning objectives produce a general-purpose backbone that supports multiple downstream sensing tasks, including human activity recognition, 3D human pose estimation, and human silhouette estimation.

%Together, these designs form a unified sensing and representation learning pipeline that transforms passively captured BFI into spatiotemporal representations of human motion. The resulting shared backbone can be paired with lightweight task-specific heads and efficiently adapted to diverse downstream sensing tasks using limited labeled data.

%We implement and evaluate BeamLens using commodity Wi-Fi devices across a variety of environments, involving different participants performing diverse human activities. Our evaluation includes three downstream sensing tasks to assess the effectiveness and generalizability of the learned representations. With the pretrained backbone frozen, BeamLens achieves an activity recognition accuracy of 90.16\%, a mean per-joint position error (MPJPE) of 8.64~cm for 3D pose estimation, and an Intersection over Union (IoU) of 70.64\% for human silhouette estimation. Additional experiments investigate the effectiveness of the learned representations under different sensing conditions and training configurations.

We implement and evaluate BeamLens using commodity Wi-Fi devices across a variety of environments, involving different participants performing diverse human activities. Our evaluation includes three downstream sensing tasks to assess the effectiveness and generalizability of the learned representations. With the pretrained backbone frozen, BeamLens achieves a human activity recognition accuracy of 90.16\%, a mean per-joint position error (MPJPE) of 8.64~cm for 3D human pose estimation, and an Intersection over Union (IoU) of 70.64\% for human silhouette estimation. Additional experiments investigate the effectiveness of the learned representations under different sensing conditions and training configurations.

Our contributions are summarized as follows:

\begin{itemize}

\item We present BeamLens, the first BFI-based Wi-Fi sensing system that exploits fine-grained spatial information to support versatile human sensing tasks, including 3D human pose estimation, human silhouette estimation, and human activity recognition.

%By exploiting the spatial information preserved in BFI, BeamLens enables diverse downstream sensing tasks through a shared representation.

\item We develop a fine-grained spatial information extraction method that jointly estimates the AoD and RToF from BFI by leveraging spatial diversity across multiple transmitting antennas and frequency diversity across multiple subcarriers.

%By employing two orthogonally oriented AP-STA links, BeamLens captures complementary azimuth and elevation views to provide more comprehensive spatial information for fine-grained Wi-Fi sensing.

%\item We develop a spatial resolution enhancement method that combines a physical sparse antenna array with a virtual antenna array to improve angular resolution and multipath separation capability.

\item We develop a spatial resolution enhancement method that combines a sparse physical antenna array with a uniform virtual antenna array to improve angular resolution and multipath separation capability.

%to enlarge the antenna aperture and improve angular resolution without adding additional antennas. By constructing a virtual antenna array based on the physical sparse antenna array, BeamLens further mitigates spatial aliasing while exploiting the enlarged aperture.

\item We develop a self-supervised representation learning framework that integrates complementary spatial information across different views and captures temporal dependencies between observations to learn a general-purpose backbone for diverse downstream human sensing tasks.

%. Cross-view contrastive learning exploits synchronized observations as natural positive pairs, while temporally weighted soft positives and masked temporal reconstruction capture temporal continuity and dependencies. Together, these learning objectives enable BeamLens to learn a general-purpose backbone that supports multiple downstream sensing tasks.

\item We implement BeamLens using commodity Wi-Fi devices and conduct extensive experiments to evaluate its performance. Results across three human sensing tasks demonstrate the effectiveness and versatility of BeamLens under various sensing conditions and training configurations.
\end{itemize}
\section{Related Work}

\subsection{CSI-Based Wi-Fi Sensing}

Wi-Fi-based wireless sensing has been extensively studied as a device-free approach to human sensing, covering activity recognition~\cite{wang2014eyes, wang2015understanding}, gesture recognition~\cite{zhang2021widar3}, indoor localization~\cite{kotaru2015spotfi, xiong2013arraytrack}, and vital-sign monitoring~\cite{liu2015tracking}. More recently, researchers have extended Wi-Fi sensing toward finer-grained human pose estimation. WiPose~\cite{jiang2020towards} is among the early systems to estimate 3D human pose from commodity Wi-Fi CSI using deep neural networks. Subsequent systems have explored free-form activity tracking, on-the-go pose estimation for moving users, and person re-identification using spatial information extracted from multi-antenna CSI measurements~\cite{ren2021winect,ren2022gopose,ren2023person}. Moreover, Person-in-Wi-Fi~\cite{wang2019person} and DensePose from Wi-Fi~\cite{geng2022densepose} show that Wi-Fi signals can support body segmentation and dense human pose correspondence when trained with vision-derived annotations, indicating that Wi-Fi measurements retain spatial information about the human body beyond coarse motion patterns.

%Despite this progress, existing Wi-Fi sensing systems rely on CSI as their underlying signal source~\cite{ma2019wifi}. CSI provides per-subcarrier amplitude and phase measurements, enabling the extraction of fine-grained spatial and temporal features for human sensing. However, obtaining CSI from commodity Wi-Fi devices remains challenging, as it is supported by a limited range of network interface cards and often requires modified drivers or firmware, such as the Linux 802.11n CSI Tool~\cite{halperin2011tool}, to expose channel measurements that are not normally accessible to users. These hardware and firmware dependencies complicate the deployment of CSI-based sensing on widely available Wi-Fi devices, motivating the exploration of alternative, more accessible sources of channel information.

Despite this progress, existing Wi-Fi sensing systems rely on CSI~\cite{ma2019wifi}, which remains difficult to obtain on commodity Wi-Fi devices. Only a small number of Wi-Fi chipsets and NICs can report CSI~\cite{wu2023enabling}, and accessing it usually relies on modified drivers, customized firmware, or dedicated extraction tools~\cite{halperin2011tool, gringoli2019free, jiang2021eliminating}.    These limitations create a significant gap between existing CSI-based systems and ubiquitous Wi-Fi sensing.

\subsection{BFI-Based Wi-Fi Sensing}

BFI has recently gained attention as a more accessible alternative to CSI. Early BFI-based systems mainly rely on coarse-grained features, such as temporal variations or statistical features of the reported feedback, to support relatively simple sensing tasks. For example, prior work has used BFI to detect human presence and movement~\cite{suraweera2019stand}, count exercise repetitions~\cite{kato2022cbr}, and recognize human activities~\cite{chen2024ship, haque2024bfa, haque2025beamsense, haque2025si}. The plaintext transmission of BFI further allows an eavesdropper to detect room occupancy through walls~\cite{xiao2025lend} and infer keystrokes for password eavesdropping~\cite{chen2024silent}.

Subsequent studies further process BFI to support a wider range of sensing tasks. Wu et al.~\cite{wu2023enabling} map bidirectional beamforming reports to the multipath channel, allowing existing CSI-based sensing applications to run on BFI. BFMSense~\cite{yi2024bfmsense} computes a BFI ratio for respiration sensing and human trajectory tracking, and FreeBFI~\cite{wang2025freebfi} combines the channel and signal-to-noise ratio (SNR) information in BFI across an arbitrary number of antennas to improve respiration monitoring and gesture recognition. Kondo et al.~\cite{kondo2022bi} exploit bidirectional beamforming feedback for human localization, Li et al.~\cite{li2024efficient} select the most informative BFI features to reduce the computational cost of positioning, and M$^2$-Fi~\cite{hu2024m} monitors the respiration of multiple people using handheld Wi-Fi devices.

Several works have also extracted spatial information from BFI. Itahara et al.~\cite{itahara2022beamforming} estimate the AoD from BFI using MUltiple SIgnal Classification (MUSIC)~\cite{schmidt1986multiple}, and BFMScan~\cite{li2026bfmscan} analyzes the subspace structure of the beamforming feedback matrix to obtain AoD profiles for angle-resolved sensing. However, both works are designed only for AoD estimation. Moreover, existing BFI-based systems are typically designed for one or a few specific tasks, and none of them supports versatile fine-grained human sensing, such as 3D human pose estimation and human silhouette estimation.

\subsection{Self-Supervised Learning for Wi-Fi Sensing}

Self-supervised learning has recently been explored to reduce the reliance of Wi-Fi sensing on large amounts of labeled data. AutoFi~\cite{yang2022autofi} exploits the geometric structure of CSI to construct self-supervised learning objectives that facilitate transfer across environments without manual annotation. Bocus et al.~\cite{bocus2022self} apply contrastive pretraining to unlabeled CSI measurements and report improved downstream activity recognition with limited labeled data. More recently, Xu et al.~\cite{xu2025evaluating} systematically evaluate self-supervised learning for CSI-based activity recognition, showing that its effectiveness depends on the choice of positive and negative pairs and on the domain gap between pretraining and downstream data. Other studies reduce the need for labeled data in new environments through domain adaptation, such as removing environment-specific features with adversarial learning~\cite{jiang2018towards} and transferring sensing models to unseen environments with only a small amount of target-domain data~\cite{zhang2018crosssense}.

Beyond Wi-Fi sensing, contrastive and masked representation learning have been widely explored for learning representations from unlabeled data. InfoNCE~\cite{oord2018representation} provides a contrastive learning objective adopted by frameworks such as SimCLR~\cite{chen2020simple}, while Contrastive Multiview Coding~\cite{tian2020contrastive} exploits agreement between different views of the same underlying observation. Subsequent frameworks improve contrastive learning with momentum encoders~\cite{he2020momentum} and pretext-invariant representations~\cite{misra2020self}, and Siamese architectures remove the need for negative pairs~\cite{grill2020bootstrap, chen2021exploring}. Cross-modal contrastive learning has further been shown to yield highly transferable representations~\cite{radford2021learning}. Masked representation learning, exemplified by MAE~\cite{he2022masked}, instead learns contextual information by reconstructing missing portions of the input, and data2vec~\cite{baevski2022data2vec} extends masked prediction into a general framework across speech, vision, and language.

Despite this progress, existing self-supervised Wi-Fi sensing methods rarely exploit complementary spatial views and the temporal continuity of human motion jointly within a single pretraining framework.
\section{Preliminary}
\label{sec:preliminary}

\subsection{Channel Sounding Procedure}
\label{sec:sounding}

\begin{figure}[t]
    \centering
    \begin{minipage}[b]{0.48\linewidth}
        \centering
        \includegraphics[width=\linewidth]{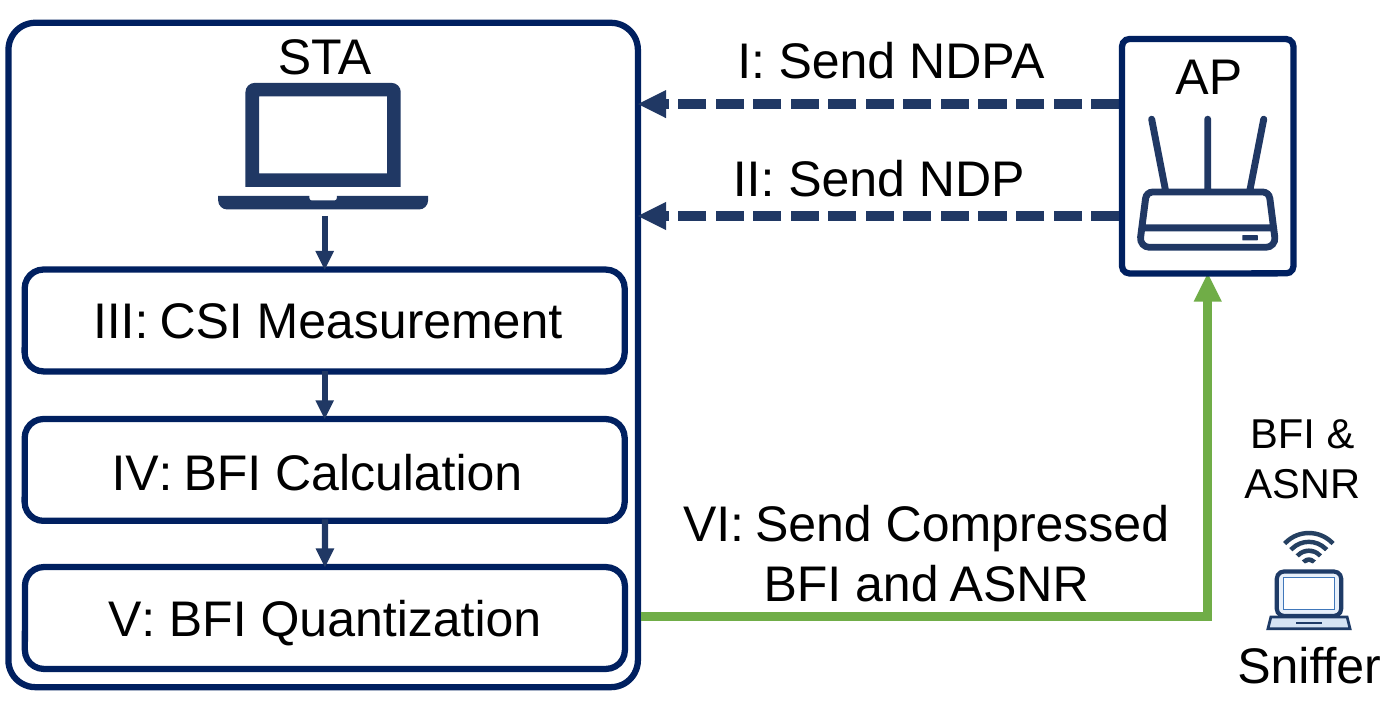}
        \caption{Channel sounding procedure. A nearby sniffer can passively capture the BFI and ASNR in the feedback frames.}
        \label{fig:bfi_process}
    \end{minipage}
    \hfill
    \begin{minipage}[b]{0.48\linewidth}
        \centering
        \includegraphics[width=\linewidth]{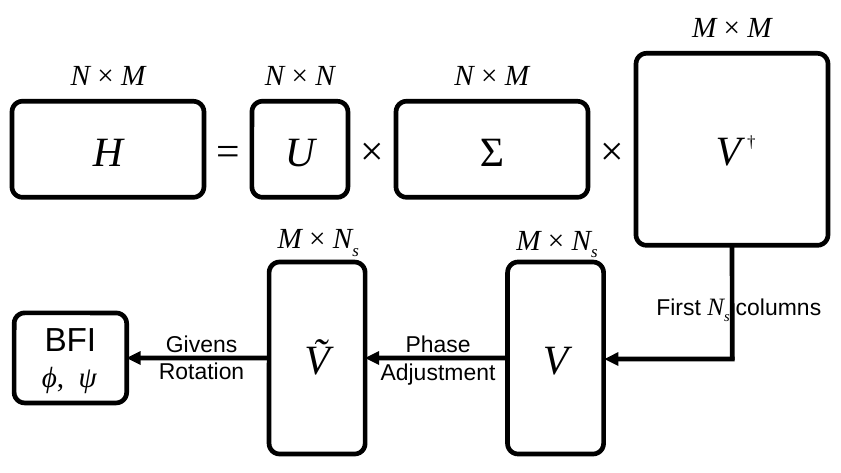}
        \caption{BFI calculation procedure at the STA, including SVD, column truncation, phase adjustment, and Givens rotation.}
        \label{fig:svd}
    \end{minipage}
\end{figure}

IEEE 802.11ac/ax~\cite{ieee80211ac2013, ieee80211ax2021} supports explicit beamforming, which enables an AP to adjust its transmit beamforming weights based on channel information reported by a STA. To obtain this information, the AP and STA perform a channel sounding procedure, as illustrated in Fig.~\ref{fig:bfi_process}. The procedure consists of the following six steps.

\textit{Step I: NDPA Transmission.} The AP initiates channel sounding by transmitting a Null Data Packet Announcement (NDPA) frame to announce the upcoming sounding procedure.

\textit{Step II: NDP Transmission.} Following the NDPA, the AP transmits a Null Data Packet (NDP) containing known training signals, allowing the STA to estimate the downlink wireless channel.

\textit{Step III: CSI Measurement.} Upon receiving the NDP, the STA estimates the downlink CSI across the reported orthogonal frequency-division multiplexing (OFDM) subcarriers. Let $\mathbf{H}_k \in \mathbb{C}^{N\times M}$ denote the CSI matrix at subcarrier $k$, where $M$ and $N$ are the numbers of AP antennas and STA antennas, respectively. Each element of $\mathbf{H}_k$ represents the complex channel response between an AP-STA antenna pair.

\textit{Step IV: BFI Calculation.} Based on the estimated CSI, the STA computes a beamforming feedback matrix that preserves the transmit-side channel information required for beamforming. Specifically, the STA performs singular value decomposition (SVD) on the CSI matrix and extracts the relevant right singular vectors to construct the feedback matrix. 

\textit{Step V: BFI Quantization.} To reduce feedback overhead, the STA compresses the feedback matrix using phase adjustments and Givens rotations. The resulting angular parameters are quantized according to IEEE 802.11ac/ax~\cite{ieee80211ac2013, ieee80211ax2021} to produce compressed BFI.

\textit{Step VI: Feedback Reporting.} The STA transmits the compressed BFI to the AP, together with the average SNR (ASNR) of each reported spatial stream. The AP uses this feedback to determine its beamforming weights for subsequent downlink transmissions. Since the feedback is transmitted without encryption, a nearby Wi-Fi sniffer can passively capture BFI and ASNR by monitoring the corresponding feedback frames, without directly accessing the CSI measurements of either device~\cite{haque2023wi, xiao2025lend}.

\subsection{BFI Calculation Procedure}
\label{sec:bfi_csi}

This section details how the STA computes BFI from the estimated CSI in Steps IV and V~\cite{wu2023enabling, yi2024bfmsense}, as illustrated in Fig.~\ref{fig:svd}.

For each reported subcarrier $k$, the STA performs SVD on the estimated CSI matrix:
\begin{equation}
    \mathbf{H}_k = \mathbf{U}_k \boldsymbol{\Sigma}_k \mathbf{V}_k^{\dagger},
    \label{eq:svd}
\end{equation}
where $\mathbf{U}_k \in \mathbb{C}^{N \times N}$ and $\mathbf{V}_k \in \mathbb{C}^{M \times M}$ contain the left and right singular vectors, respectively, $\mathbf{\Sigma}_k \in \mathbb{R}^{N \times M}$ contains the nonnegative singular values, and $(\cdot)^\dagger$ denotes the conjugate transpose. The right singular vectors in $\mathbf{V}_k$ characterize the transmit-side spatial structure of the wireless channel.

Rather than reporting the complete matrix $\mathbf{V}_k$, the STA retains its first $N_s$ columns, where $N_s \leq \min(M,N)$ is the number of reported spatial streams. For simplicity, we continue to denote the retained matrix by $\mathbf{V}_k \in \mathbb{C}^{M\times N_s}$. The STA then performs column-wise phase adjustment by subtracting the phase of the last element in each column. The resulting matrix can be expressed as
\begin{equation}
    \widetilde{\mathbf{V}}_k = \mathbf{V}_k\mathbf{D}_k,
    \label{eq:phase_norm}
\end{equation}
where $\mathbf{D}_k\in\mathbb{C}^{N_s\times N_s}$ is a diagonal matrix with unit-modulus entries that removes the reference phase of each column while preserving the relative phase relationships among its elements. The resulting matrix $\widetilde{\mathbf{V}}_k$ is further parameterized using phase angles $\phi$ and rotation angles $\psi$ through Givens rotations. These angles are quantized and transmitted as compressed BFI.
\section{System Design}
\label{sec:design}

\subsection{System Overview}
\label{sec:overview}

\begin{figure}[t]
    \centering
    \includegraphics[width=\linewidth]{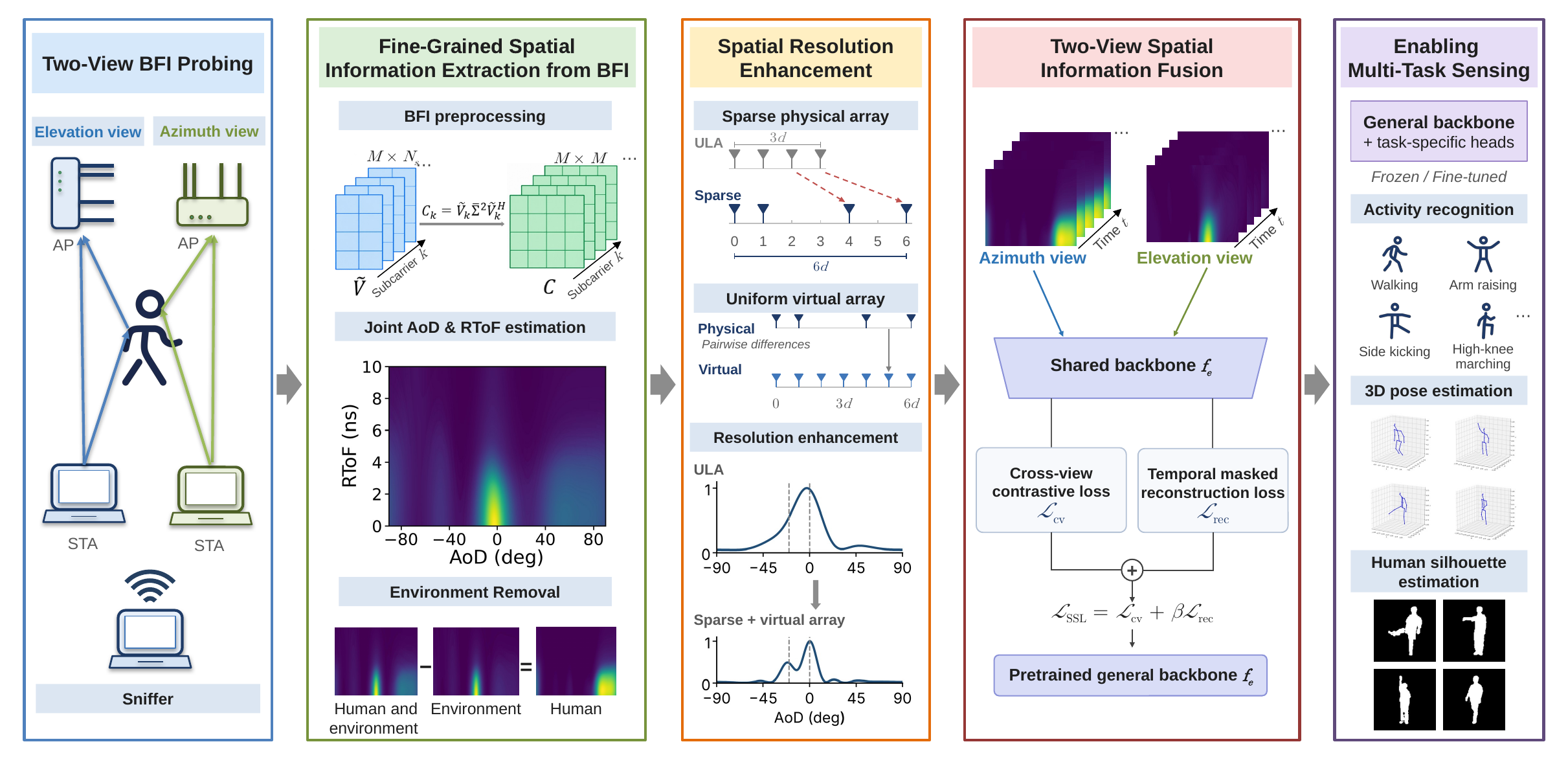}
    \caption{Overview of BeamLens. Two orthogonally oriented AP-STA links provide the azimuth and elevation views. For each view, BeamLens extracts AoD-RToF heatmaps from the captured BFI, enhances their spatial resolution with a sparse physical array and a uniform virtual array, and fuses the two views through self-supervised learning to enable versatile human sensing.}
    \label{fig:overview}
\end{figure}

As shown in Fig.~\ref{fig:overview}, BeamLens consists of four modules: two-view BFI probing, fine-grained spatial information extraction, spatial resolution enhancement, and two-view spatial information fusion. In the two-view BFI probing module, BeamLens deploys two AP-STA links, each consisting of a commodity Wi-Fi AP and a STA. The transmitting antenna arrays of the two APs are oriented horizontally and vertically, providing an azimuth view and an elevation view of the human body, respectively. While a person moves between the two links, each AP periodically performs channel sounding with its STA, and a nearby sniffer passively captures the BFI and ASNR reported by both STAs.

In the fine-grained spatial information extraction module, BeamLens first preprocesses the captured BFI and ASNR of each view into a matrix that preserves the phase differences across transmitting antennas and subcarriers. It then jointly estimates the AoD and RToF of the propagation paths, producing an AoD-RToF heatmap for each sensing interval. Since these heatmaps also contain reflections from the environment, such as the line-of-sight (LoS) path, walls, and furniture, BeamLens further removes these environment-dependent components and retains the reflections from the human body. In the spatial resolution enhancement module, BeamLens rearranges the transmitting antennas of each AP into a sparse physical array to enlarge the antenna array aperture. Note that constructing the sparse physical antenna array only requires rearranging the existing antennas of a Wi-Fi device. It does not introduce any additional hardware. Based on the sparse physical array, BeamLens further constructs a uniform virtual array to increase the number of propagation paths that can be resolved.

Finally, in the two-view spatial information fusion module, BeamLens learns a shared backbone from unlabeled sequences of azimuth and elevation heatmaps through temporal-aware cross-view contrastive learning and temporal masked reconstruction. The pretrained backbone is then combined with lightweight task-specific heads, enabling versatile human sensing, including human activity recognition, 3D human pose estimation, and human silhouette estimation.

\subsection{Fine-Grained Spatial Information Extraction}
\label{sec:spatial_processing}

\subsubsection{BFI Preprocessing}
\label{sec:bfi_preprocessing}

For each BFI report, the sniffer obtains the phase-adjusted right singular vectors $\widetilde{\mathbf V}_k$ of each reported subcarrier $k$, together with the ASNR of each spatial stream. BeamLens first converts the ASNR into a diagonal weighting matrix~\cite{itahara2022beamforming}:
\begin{equation}
    \overline{\mathbf\Sigma}^2
    =
    \operatorname{diag}\left(
    10^{\mathrm{ASNR}_1/10},
    \ldots,
    10^{\mathrm{ASNR}_{N_s}/10}
    \right),
    \label{eq:asnr_weight}
\end{equation}
where $\mathrm{ASNR}_i$ is the ASNR of the $i$-th stream in dB. Since ASNR is averaged over subcarriers, $\overline{\mathbf\Sigma}^2$ provides the relative strengths of the streams and approximates their squared singular values. BeamLens then computes
\begin{equation}
    \mathbf C_k
    =
    \widetilde{\mathbf V}_k
    \overline{\mathbf\Sigma}^2
    \widetilde{\mathbf V}_k^\dagger .
    \label{eq:corr}
\end{equation}
Substituting Eq.~\eqref{eq:phase_norm} into Eq.~\eqref{eq:corr}, the phase adjustment cancels out since $\mathbf D_k\overline{\mathbf\Sigma}^2\mathbf D_k^\dagger=\overline{\mathbf\Sigma}^2$, which yields $\mathbf C_k = \mathbf V_k\overline{\mathbf\Sigma}^2\mathbf V_k^\dagger$. Therefore, $\mathbf C_k$ can be compared across subcarriers. Moreover, according to Eq.~\eqref{eq:svd},
\begin{equation}
    \mathbf H_k^\dagger\mathbf H_k
    =
    \mathbf V_k
    \mathbf\Sigma_k^\dagger\mathbf\Sigma_k
    \mathbf V_k^\dagger ,
    \label{eq:channel_corr}
\end{equation}
where $\mathbf{V}_k$ denotes the complete right singular vector matrix in Eq.~\eqref{eq:svd}, rather than the truncated matrix used for BFI reporting. Comparing Eqs.~\eqref{eq:corr} and~\eqref{eq:channel_corr}, $\mathbf C_k$ approximates $\mathbf H_k^\dagger\mathbf H_k$, which contains the phase differences of the channel across transmitting antennas and subcarriers. BeamLens therefore uses $\mathbf C_k$ for the subsequent joint AoD and RToF estimation.

\subsubsection{Joint AoD and RToF Estimation}
\label{sec:heatmap_construction}

\begin{figure}[t]
    \centering
    \includegraphics[width=0.45\columnwidth]{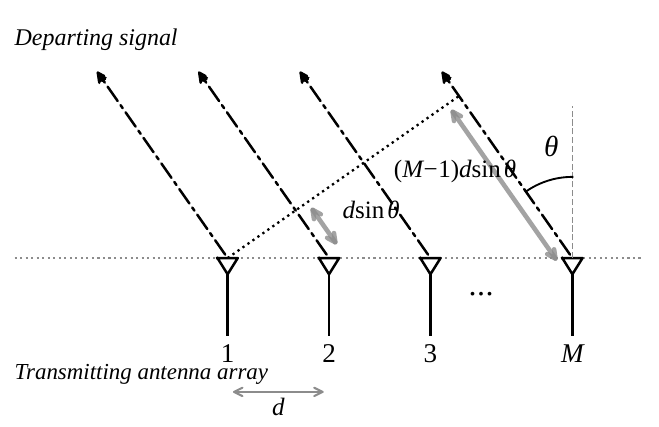}
    \caption{Phase shifts across transmitting antennas for a path departing at AoD $\theta$.}
    \label{fig:spatial_info}
\end{figure}

\begin{figure*}[t]
    \centering
    \includegraphics[width=\textwidth]{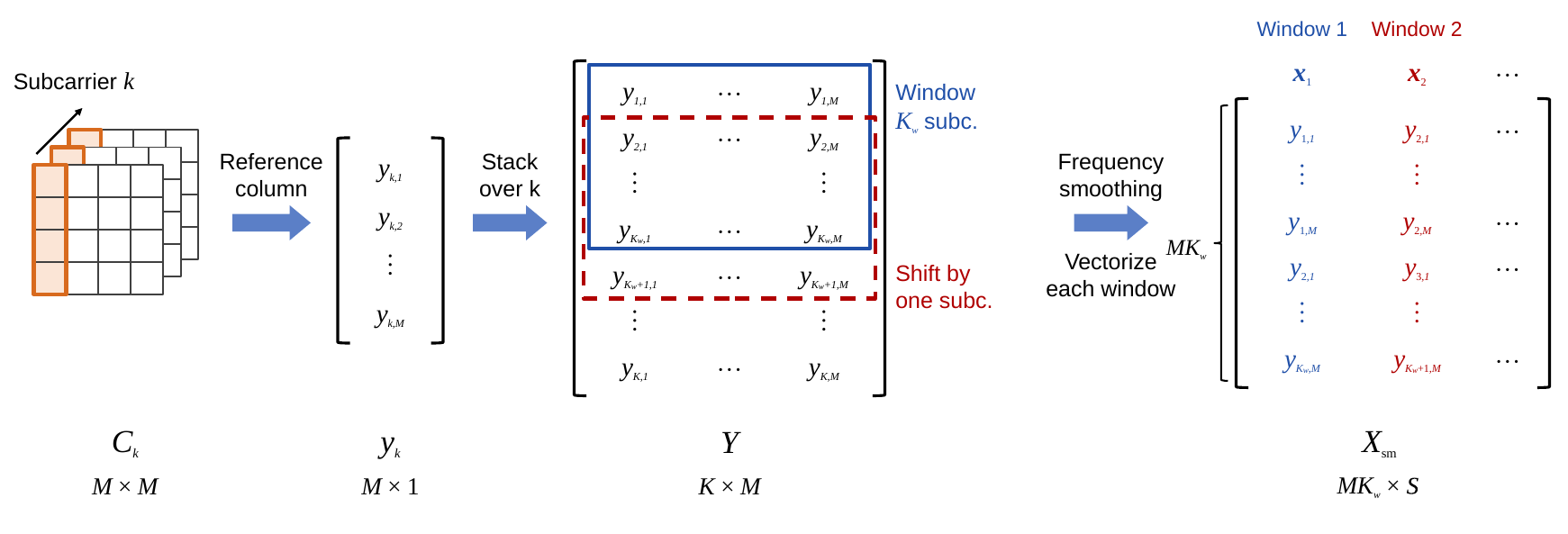}
    \caption{Construction of the smoothed measurement matrix via frequency smoothing.}
    \label{fig:smoothing}
\end{figure*}

A propagation path changes the phase of the signal in two ways. First, its AoD $\theta$ makes the phase differ across transmitting antennas. As shown in Fig.~\ref{fig:spatial_info}, the $m$-th antenna, located at $p_m d$ from the first one, has an extra phase of $p_m\mu(\theta)$, where $p_m$ is its normalized position with $p_1=0$, $\mu(\theta)=2\pi d\sin\theta/\lambda$, $\lambda$ is the wavelength, and $d=\lambda/2$ is the unit antenna spacing. Second, its delay $\tau$ makes the phase differ across subcarriers: subcarrier $k$ with frequency $f_k$ has a phase of $-2\pi f_k\tau$. The channel is the sum of all paths:
\begin{equation}
    \mathbf H_k
    =
    \sum_{l=0}^{L}
    \alpha_l e^{-j2\pi f_k\tau_l}
    \mathbf b_l\mathbf a^T(\theta_l),
    \quad
    \mathbf a(\theta)
    =
    \left[
        e^{jp_1\mu(\theta)},
        \ldots,
        e^{jp_M\mu(\theta)}
    \right]^T,
    \label{eq:channel_model}
\end{equation}
where $\alpha_l$, $\theta_l$, and $\tau_l$ are the gain, AoD, and delay of the $l$-th path, $\mathbf b_l$ is its response at the receiving antennas, $l=0$ denotes the LoS path, and $(\cdot)^{T}$ denotes the transpose. The vector $\mathbf a(\theta)$, known as the steering vector~\cite{van2002optimum}, collects the phases across transmitting antennas.

Substituting Eq.~\eqref{eq:channel_model} into $\mathbf H_k^\dagger\mathbf H_k$ and using $p_1=0$, the first column becomes
\begin{equation}
    [\mathbf H_k^\dagger\mathbf H_k]_{:,1}
    =
    \sum_{l=0}^{L}\sum_{l'=0}^{L}
    \alpha_l^*\alpha_{l'}\,(\mathbf b_l^\dagger\mathbf b_{l'})\,
    e^{j2\pi f_k(\tau_l-\tau_{l'})}\,
    \mathbf a^*(\theta_l),
    \label{eq:gram_column}
\end{equation}
in which the absolute delays cancel and only delay differences remain, and $(\cdot)^{*}$ denotes the complex conjugate. BeamLens takes the first column of $\mathbf C_k$. Since the LoS path is usually the strongest indoors, the products involving the LoS path dominate. Among them, the terms with $l=0$ and $l'\neq0$ lie at negative RToF. These terms remain in the measurement and may still affect the subspace estimation through the finite frequency aperture, but their main contributions lie outside the nonnegative RToF range that BeamLens searches. Under this dominant-LoS cross-term approximation, the column is modeled for the nonnegative RToF search by the terms with $l'=0$ in Eq.~\eqref{eq:gram_column}:
\begin{equation}
    \mathbf y_k
    =
    [\mathbf C_k]_{:,1}
    \approx
    \sum_{l=0}^{L}
    \gamma_l\,
    e^{j2\pi f_k\Delta\tau_l}\,
    \mathbf a^*(\theta_l),
    \label{eq:reference_column}
\end{equation}
where $\gamma_l=\alpha_l^*\alpha_0\,\mathbf b_l^\dagger\mathbf b_0$ is the complex gain of the $l$-th term and $\Delta\tau_l=\tau_l-\tau_0$ is the RToF of the $l$-th path with respect to the LoS path, so that the $l=0$ term is the LoS self-product at zero RToF. Each path thus contributes a conjugated steering vector determined by its AoD, whose phase rotates across subcarriers at a rate determined by its RToF. The products between two reflected paths, including their self-products, are much weaker and are neglected. When the LoS path is blocked, the strongest path serves as the reference instead, and RToF is measured relative to it; paths arriving earlier than this reference then fall at negative RToF and are not captured, and the neglected products between strong reflected paths may become non-negligible. Eq.~\eqref{eq:reference_column} has the same form as the CSI model in SpotFi~\cite{kotaru2015spotfi}, which allows BeamLens to adopt its super-resolution approach.

Following SpotFi, BeamLens treats every antenna-subcarrier pair as a sensor. For each BFI report $b$, it stacks the reference columns of all $K$ reported subcarriers into
\begin{equation}
    \mathbf Y^{(b)}
    =
    \left[
        \mathbf y_1^{(b)},
        \ldots,
        \mathbf y_K^{(b)}
    \right]^T
    \in\mathbb C^{K\times M},
    \label{eq:stacked_obs}
\end{equation}
in which the phase variation across antennas within each row encodes the AoD, while the phase variation across subcarriers along each column encodes the RToF. Since the paths in a single observation are coherent, MUSIC~\cite{schmidt1986multiple} cannot separate them directly. BeamLens therefore applies frequency smoothing~\cite{shan1985spatial, kotaru2015spotfi}, as shown in Fig.~\ref{fig:smoothing}: a window of $K_w$ consecutive subcarriers slides along $\mathbf Y^{(b)}$ one subcarrier at a time, and each window is vectorized into $\mathbf x_s^{(b)}\in\mathbb C^{MK_w}$. Shifting the window only changes the phase of each path's gain, so different windows provide different phase combinations of the same paths, which improves the effective rank of the covariance matrix and facilitates the separation of coherent paths. Since the reported subcarriers are not uniformly spaced around the null subcarriers at the band center, and the frequency steering vector below assumes uniform spacing, only windows whose subcarriers are uniformly spaced are used, and we denote their number by $S$. This yields the smoothed measurement matrix
\begin{equation}
    \mathbf X_{\mathrm{sm}}^{(b)}
    =
    \begin{bmatrix}
        \mathbf x_1^{(b)} &
        \cdots &
        \mathbf x_S^{(b)}
    \end{bmatrix}
    \in\mathbb C^{MK_w\times S}.
    \label{eq:smoothed_measurements}
\end{equation}
BeamLens then averages over the BFI reports $\mathcal B_t$ within each sensing interval $t$ to obtain the covariance matrix
\begin{equation}
    \mathbf R_t
    =
    \frac{1}{|\mathcal B_t|S}
    \sum_{b\in\mathcal B_t}
    \mathbf X_{\mathrm{sm}}^{(b)}
    \left(\mathbf X_{\mathrm{sm}}^{(b)}\right)^\dagger .
    \label{eq:heatmap_covariance}
\end{equation}
In our implementation, each sensing interval spans 1\,s (about 30 BFI reports) and slides by one report, so the resulting heatmap sequence keeps the 30\,Hz report rate.

Finally, BeamLens applies MUSIC to $\mathbf R_t$. According to Eq.~\eqref{eq:reference_column}, the joint steering vector of a path with AoD $\theta$ and RToF $\tau$ over a window is
\begin{equation}
    \mathbf a(\theta,\tau)
    =
    \mathbf a_f(\tau)\otimes\mathbf a^*(\theta),
    \quad
    \mathbf a_f(\tau)
    =
    \left[
        1,
        e^{j2\pi\Delta f\tau},
        \ldots,
        e^{j2\pi(K_w-1)\Delta f\tau}
    \right]^T,
    \label{eq:joint_steering}
\end{equation}
where $\Delta f=N_g\Delta f_{\mathrm{sc}}$ is the spacing between adjacent reported subcarriers, $N_g$ is the subcarrier grouping factor, $\Delta f_{\mathrm{sc}}$ is the OFDM subcarrier spacing, and $\otimes$ denotes the Kronecker product. BeamLens performs eigenvalue decomposition of $\mathbf R_t$ and takes the eigenvectors of the $D$ largest eigenvalues as the signal subspace, where $D$ is the smallest number of eigenvalues whose sum exceeds a fraction $\epsilon$ of the total, with $\epsilon$ chosen empirically. The remaining eigenvectors form the noise subspace $\mathbf E_{\mathrm n,t}$, and the MUSIC pseudospectrum is
\begin{equation}
    P_t(\theta,\tau)
    =
    \frac{1}{
        \mathbf a^\dagger(\theta,\tau)
        \mathbf E_{\mathrm n,t}
        \mathbf E_{\mathrm n,t}^\dagger
        \mathbf a(\theta,\tau)
    }.
    \label{eq:angle_delay_music}
\end{equation}
$P_t(\theta,\tau)$ peaks at the AoD and RToF of each path. Evaluating it over predefined AoD and nonnegative RToF grids yields an AoD-RToF heatmap, and repeating this over successive sensing intervals produces a heatmap sequence. Note that the nominal delay resolution of the 80\,MHz channel, about 12.5\,ns, applies to Fourier-based estimation. As a subspace method, MUSIC achieves super-resolution whose accuracy depends on the SNR and the number of snapshots, and each $\mathbf R_t$ aggregates $|\mathcal B_t|S$ smoothed snapshots. Moreover, BeamLens does not rely on RToF alone to separate paths: paths are resolved jointly in AoD and RToF, and the downstream networks exploit the continuous AoD-RToF patterns rather than individual peak locations.

\subsubsection{Environment Removal}
\label{sec:environment_removal}

\begin{figure*}[t]
\centering
\includegraphics[width=\textwidth]{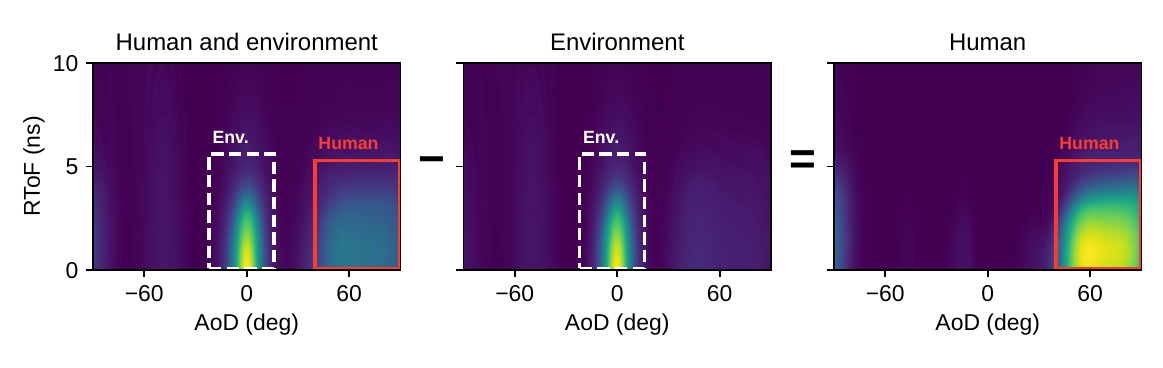}
\caption{Environment removal.}
\label{fig:environment_removal}
\end{figure*}

We denote the heatmap of sensing view $v$ in interval $t$ as $\mathbf{S}_t^{(v)}$. Besides reflections from the human body, $\mathbf{S}_t^{(v)}$ also contains reflections from the environment, including the LoS path, walls, furniture, and other static objects, which vary across environments. To make the heatmaps focus on the human body, BeamLens removes these environment-dependent components and retains the reflections from the human body.

Specifically, BeamLens applies background subtraction to remove the environmental reflections. Before each sensing session, it collects several seconds of BFI in the empty environment, processes them with the same pipeline, and averages the resulting heatmaps into a background reference $\mathbf S_{\mathrm{bg}}^{(v)}$ for each view. The same empty-environment BFI is also used to calibrate the per-antenna phase offsets introduced by the transmit chains and the radio-frequency (RF) extension cables~\cite{gjengset2014phaser}. Since the LoS path dominates in the empty environment and its AoD $\theta_0$ is known from the deployment geometry, BeamLens estimates these offsets by comparing the phases of $\mathbf y_k$ with $\mathbf a^*(\theta_0)$ and compensates them in all subsequent measurements. Since the overall magnitude of the heatmaps may differ between the two measurements, both are normalized before subtraction:
\begin{equation}
\widetilde{\mathbf S}_t^{(v)}
=
\max\!\left(
\mathrm{Norm}\big(\mathbf S_t^{(v)}\big)
-
\mathrm{Norm}\big(\mathbf S_{\mathrm{bg}}^{(v)}\big),
\,0
\right),
\label{eq:background_subtraction}
\end{equation}
where $\mathrm{Norm}(\cdot)$ rescales a heatmap to $[0,1]$ by min-max normalization and the maximum is taken element-wise. As shown in Fig.~\ref{fig:environment_removal}, this suppresses the dominant static components while highlighting the changes caused by the human. It is a background suppression step rather than an exact separation of human reflections: components weakened by the human are discarded by the truncation, and the per-heatmap normalization removes absolute energy differences between frames. The resulting heatmap sequences $\{\widetilde{\mathbf S}_t^{(v)}\}$ serve as the inputs for representation learning.

\subsection{Spatial Resolution Enhancement}
\label{sec:spatial_enhancement}

%The spatial resolution of the AoD-RToF heatmaps is limited by the few transmitting antennas of commodity APs in two ways. First, the small aperture makes it difficult to distinguish paths with similar AoDs. Second, the limited number of physical antennas restricts the available spatial degrees of freedom, which can limit the separation of multiple propagation paths. BeamLens addresses both limitations with the same antennas: it rearranges the antennas into a sparse layout to enlarge the aperture, and then constructs a virtual array with more elements than the physical array to enlarge the subspace available to MUSIC.

\subsubsection{Sparse Physical Antenna Array Construction}
\label{sec:sparse_transmit_antenna_design}

The angular resolution of an antenna array depends on its aperture, i.e., the distance between the two outermost antennas. A uniform linear array (ULA) with $M$ antennas spaced by $d$ has an aperture of only $(M-1)d$. Simply increasing the spacing enlarges the aperture but causes ambiguity: when two adjacent antennas are more than $\lambda/2$ apart, paths from different directions can produce the same phases across the antennas, resulting in false peaks in the spectrum.

\begin{figure}[t]
\centering
\includegraphics[width=0.5\linewidth]{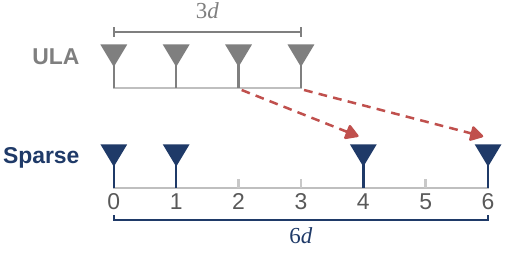}
\caption{Construction of the sparse physical antenna array from a four-antenna ULA. Moving the antennas at positions 2 and 3 to positions 4 and 6 yields the layout $\{0,1,4,6\}d$, which enlarges the aperture from $3d$ to $6d$.}
\label{fig:sparse_array}
\end{figure}

BeamLens instead constructs a sparse physical antenna array by placing the antennas at nonuniform positions, keeping some antennas $d$ apart while moving others farther away. The closely spaced antennas mitigate spatial aliasing, and the distant ones enlarge the aperture. Specifically, BeamLens uses the layout $\{p_m\}=\{0,1,4,6\}$, i.e., the antennas are placed at $\{0,1,4,6\}d$, as shown in Fig.~\ref{fig:sparse_array}. The distances between its antenna pairs cover every value from $d$ to $6d$, so the spacing $d$ is still present to mitigate spatial aliasing, while the aperture is enlarged from $3d$ to $6d$. This layout is a minimum redundancy array (MRA)~\cite{moffet1968minimum}, i.e., a configuration whose pairwise distances continuously cover $d$ to the aperture with as few repeated distances as possible, giving a larger aperture than a ULA with the same number of antennas, as listed in Table~\ref{tab:mra}. The complete set of distances is also used to construct the uniform virtual antenna array below.

\begin{table}[t]
\centering
\caption{Comparison of ULA and MRA configurations for different numbers of transmitting antennas. All antenna positions and apertures are normalized by $d$.}
\label{tab:mra}
\begin{tabular}{c c c c c}
\hline
$M$ & ULA Positions & ULA Aperture & MRA Positions & MRA Aperture \\
\hline
3 & $\{0,1,2\}$ & 2 & $\{0,1,3\}$ & 3 \\
4 & $\{0,1,2,3\}$ & 3 & $\{0,1,4,6\}$ & 6 \\
5 & $\{0,1,2,3,4\}$ & 4 & $\{0,1,4,7,9\}$ & 9 \\
6 & $\{0,1,2,3,4,5\}$ & 5 & $\{0,1,2,6,10,13\}$ & 13 \\
\hline
\end{tabular}
\end{table}

\subsubsection{Uniform Virtual Antenna Array Construction}
\label{sec:virtual_array}

\begin{figure*}[t]
\centering
\includegraphics[width=\textwidth]{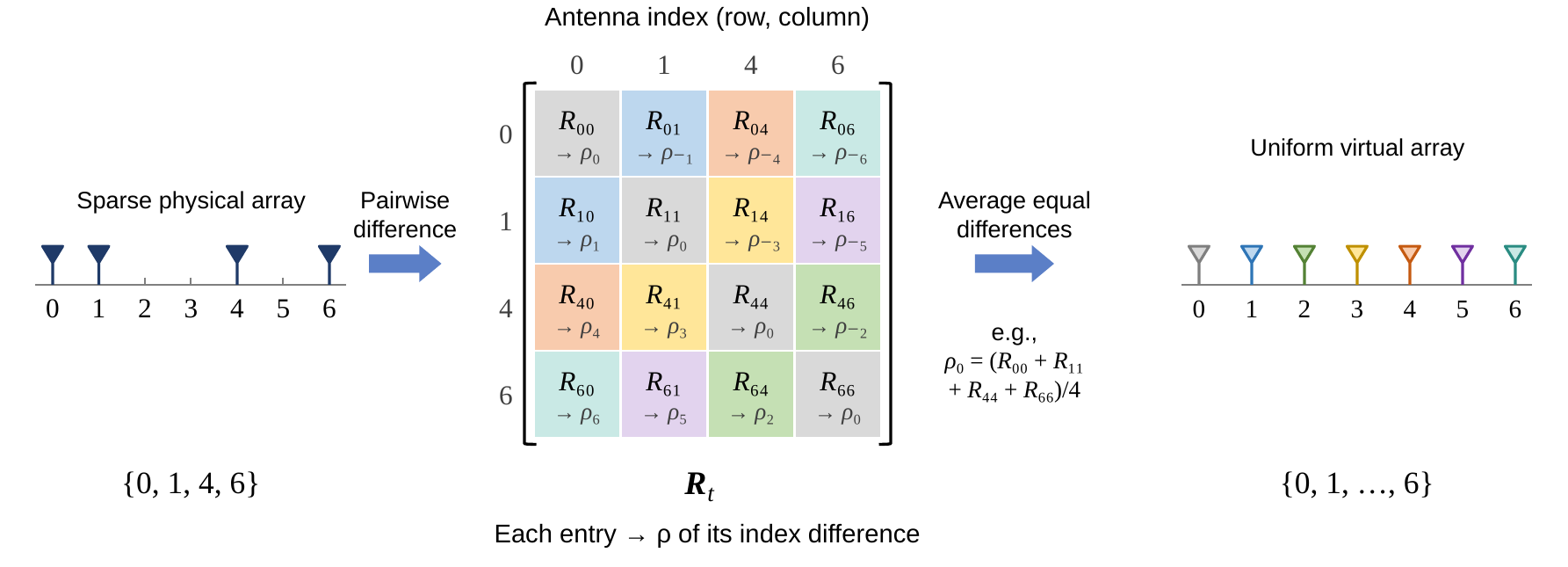}
\caption{Uniform virtual antenna array construction from the constructed sparse physical antenna array at $\{0,1,4,6\}d$. Covariance entries with identical antenna position differences (same color) are averaged, yielding a seven-element uniform virtual array at $\{0,1,\dots,6\}d$. Only the spatial dimension is shown.}
\label{fig:virtual_array}
\end{figure*}

The sparse physical array enlarges the aperture, but $\mathbf R_t\in\mathbb C^{MK_w\times MK_w}$ still has only $M=4$ antennas in its spatial dimension, which limits the number of paths that MUSIC can separate in the angular domain. Based on the sparse physical array, BeamLens therefore constructs a uniform virtual antenna array from the pairwise separations of its antennas~\cite{pillai1985new, pal2010nested}. When the paths are uncorrelated, the correlation between two antennas depends only on their separation rather than on their absolute positions. Frequency smoothing and averaging over multiple reports reduce the correlation between paths, so this property approximately holds. As shown in Fig.~\ref{fig:virtual_array}, the pairwise separations of $\{0,1,4,6\}d$ cover thirteen consecutive lags $\{-6,\dots,6\}d$, which are exactly the lags of a uniform virtual array with $M_\mathrm{v}=p_M+1=7$ elements at $\{0,1,\dots,6\}d$.

Let $[\mathbf R_t]_{(q,m),(q',m')}$ denote the entry of $\mathbf R_t$ associated with antenna $m$ at subcarrier offset $q$ and antenna $m'$ at subcarrier offset $q'$ within a window. When the paths are decorrelated and the subcarriers within a window are uniformly spaced, each entry depends only on the antenna separation $p_m-p_{m'}$ and the subcarrier offset difference $q-q'$, rather than on the absolute antenna positions and subcarrier indices. Since $\mathbf R_t$ is built from the BFI-derived correlation matrices rather than from received array snapshots, and is affected by the ASNR approximation and finite averaging, this structure holds only approximately; it relies on the approximate path model in Eq.~\eqref{eq:reference_column} and on frequency smoothing to reduce the influence of cross-terms between paths. Under this assumption, BeamLens averages all entries that share the same antenna separation $\kappa$ and subcarrier offset difference $\eta$:
\begin{equation}
\rho_t(\kappa,\eta)=
\frac{1}{|\mathcal S_{\kappa,\eta}|}
\sum_{(q,m,q',m')\in\mathcal S_{\kappa,\eta}}
[\mathbf R_t]_{(q,m),(q',m')},
\label{eq:virtual_lag}
\end{equation}
where $\mathcal S_{\kappa,\eta}=\{(q,m,q',m')\mid p_m-p_{m'}=\kappa,\ q-q'=\eta\}$. The averaged values are then arranged into the covariance matrix of the uniform virtual array:
\begin{equation}
[\bar{\mathbf R}_t]_{(q,u),(q',u')}=\rho_t(u-u',\,q-q'),
\label{eq:virtual_cov}
\end{equation}
where $u,u'\in\{0,\ldots,M_{\mathrm v}-1\}$ index the virtual antennas and $q,q'\in\{0,\ldots,K_w-1\}$ index the subcarriers within the window. The spatial dimension of $\bar{\mathbf R}_t\in\mathbb C^{M_{\mathrm v}K_w\times M_{\mathrm v}K_w}$ increases from $M=4$ to $M_{\mathrm v}=7$, while the frequency dimension $K_w$ is unchanged, providing an enlarged virtual spatial dimension for subspace estimation. When the paths are sufficiently decorrelated, this allows more paths to be separated in the angular domain; paths with nearly identical RToF may remain correlated after frequency smoothing and are harder to separate. Since lag averaging does not guarantee a valid covariance matrix, $\bar{\mathbf R}_t$ is projected onto the positive semidefinite cone by setting its negative eigenvalues to zero before subspace selection.

Finally, BeamLens computes the MUSIC pseudospectrum in Eq.~\eqref{eq:angle_delay_music} on $\bar{\mathbf R}_t$, using the joint steering vector $\mathbf a_f(\tau)\otimes\bar{\mathbf a}^*(\theta)$ with the steering vector of the uniform virtual array
\begin{equation}
\bar{\mathbf a}(\theta)=\left[1,e^{j\mu(\theta)},\ldots,e^{j(M_{\mathrm v}-1)\mu(\theta)}\right]^T.
\label{eq:virtual_steering}
\end{equation}
Since the frequency dimension is preserved, RToF is still estimated jointly with AoD.

\begin{figure}[t]
\centering
\includegraphics[width=\columnwidth]{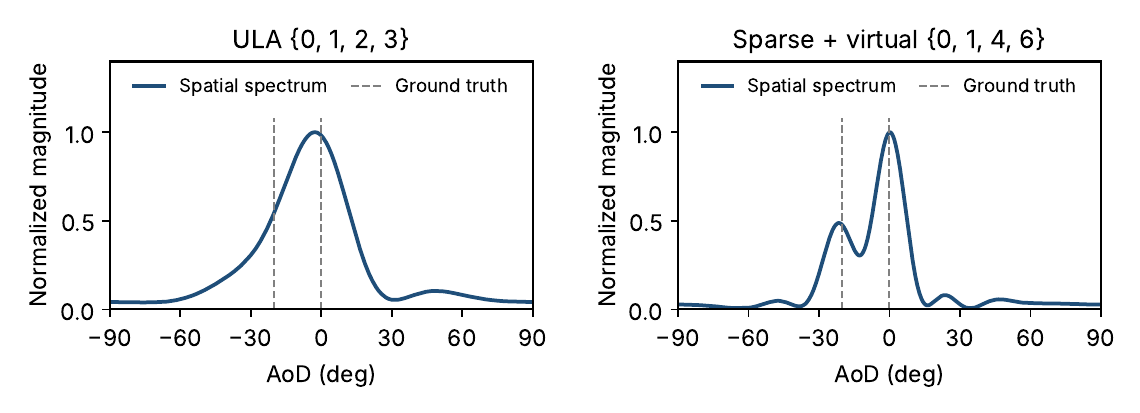}
\caption{Comparison of AoD spectra between the conventional ULA and the proposed sparse physical antenna array with uniform virtual array construction. The dashed lines indicate the ground-truth AoDs of two closely spaced paths. The ULA merges the two paths into a single broad peak, whereas the proposed design resolves them as two separate peaks.}
\label{fig:aod_enhancement}
\end{figure}

Fig.~\ref{fig:aod_enhancement} compares the AoD spectra of two closely spaced paths obtained with four antennas arranged as a ULA and as the proposed design. The ULA merges the two paths into a single broad peak, whereas the proposed design produces a narrower main lobe and clearly separates the two paths.

\subsection{Two-View Spatial Information Fusion}
\label{sec:two_view_fusion}

\begin{figure}[!htbp]
    \centering
    \includegraphics[width=0.65\linewidth]{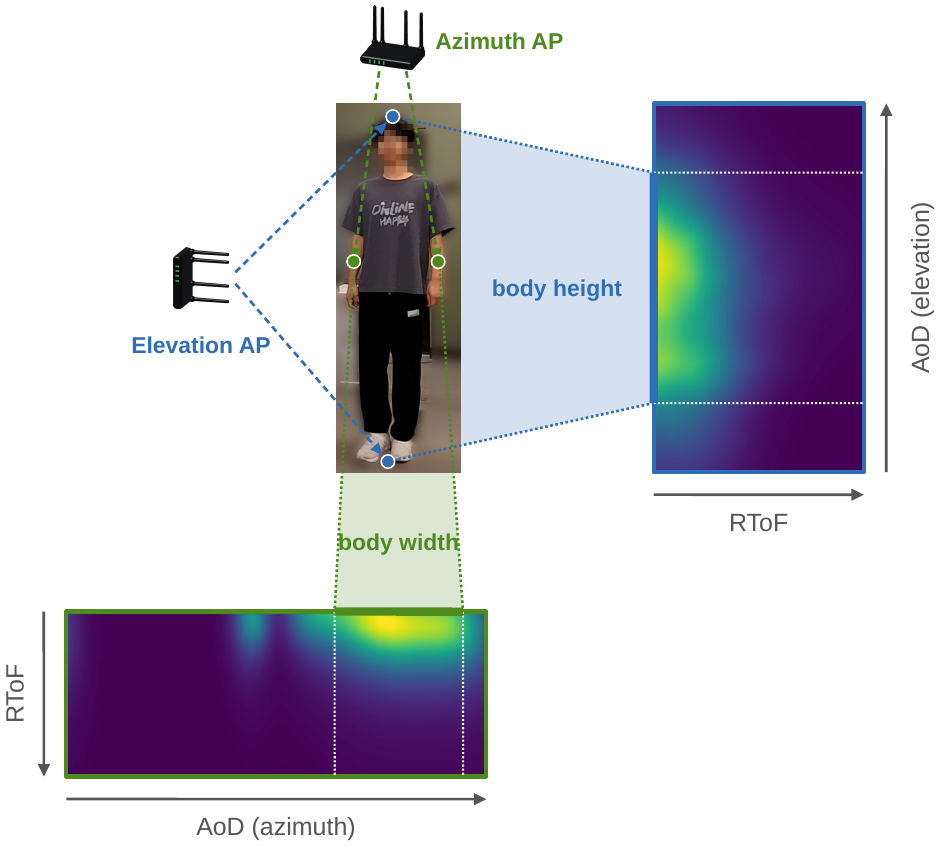}
    \caption{Two-view sensing with horizontally (azimuth) and vertically (elevation) oriented AP antenna arrays. The azimuth view resolves the horizontal extent of the body, and the elevation view resolves its vertical extent.}
    \label{fig:two_view}
\end{figure}

The two views resolve the propagation paths along different spatial dimensions. As illustrated in Fig.~\ref{fig:two_view}, the elevation view captures the vertical extent of the body, while the azimuth view captures its horizontal extent. Successive heatmaps further show how these paths change as a person moves. Neither view alone fully describes the motion, and labeling enough data for every sensing task is costly. BeamLens therefore learns a shared backbone from unlabeled heatmap sequences by aligning the two views and by recovering missing observations from their temporal context, and then reuses it for all downstream tasks.

\begin{figure*}[t]
    \centering
    \includegraphics[width=\textwidth]{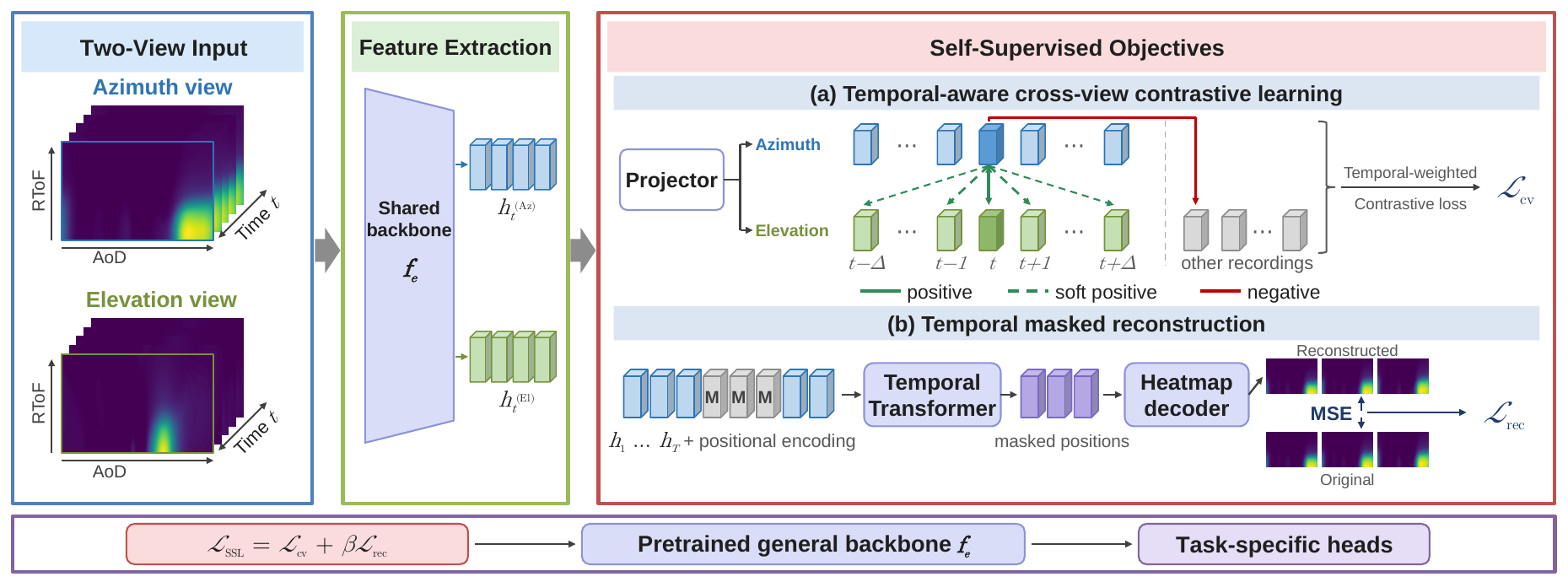}
    \caption{Self-supervised two-view spatial information fusion. The shared backbone encodes the azimuth and elevation heatmap sequences and is trained with temporal-aware cross-view contrastive learning and temporal masked reconstruction. After pretraining, only the shared backbone is kept for the downstream tasks.}
    \label{fig:ssl}
\end{figure*}

\subsubsection{Temporal-Aware Cross-View Contrastive Learning}
\label{sec:contrastive}

The azimuth and elevation views capture the same human motion at the same time, and their frames are aligned by capture timestamps. Since the reports of both links are captured by the same sniffer, they share a common clock; with both links sounding at 30\,Hz, the residual offset between paired frames is below 17\,ms, negligible compared with the 1\,s aggregation interval. Their heatmaps look different, but the changes in both are caused by the same body movement. This correspondence comes from the deployment itself and needs no labels, so BeamLens uses it to fuse the two views, in the spirit of contrastive multiview learning~\cite{tian2020contrastive}.

As shown in Fig.~\ref{fig:ssl}, BeamLens first extracts a frame-level feature from each background-removed heatmap with a shared backbone $f_e$:
\begin{equation}
    \mathbf h_t^{(v)} = f_{e}\big(\widetilde{\mathbf S}_t^{(v)}\big).
    \label{eq:ssl_feature}
\end{equation}
Sharing the backbone places both views in a common feature space. Each feature is then mapped by a lightweight projector $g_{\mathrm p}$~\cite{chen2020simple} and normalized to unit length:
\begin{equation}
    \mathbf z_t^{(v)} = \frac{g_{\mathrm p}\big(\mathbf h_t^{(v)}\big)}{\big\|g_{\mathrm p}\big(\mathbf h_t^{(v)}\big)\big\|_2},
    \label{eq:proj}
\end{equation}
so that the inner product equals the cosine similarity.

An azimuth-view frame at time $t$ serves as an anchor, and the elevation-view frame captured at the same time is its positive sample. This encourages the backbone to capture motion-related information shared across the two views, while the view-specific features remain available for downstream fusion. Since human motion changes gradually, frames close in time show nearly the same body state and should not be treated as negatives. BeamLens therefore treats the elevation-view frames of the same recording within $[t-\Delta,t+\Delta]$ as soft positives, similar to time-contrastive learning~\cite{sermanet2018time}, whose weights decrease as the time gap grows. The remaining elevation-view frames in the mini-batch, from other recordings or distant moments, are treated as unmatched observations, i.e., negatives. This definition is based on temporal correspondence rather than activity labels, so some negatives may contain similar postures.

Formally, a mini-batch contains $B$ synchronized frame pairs, where the $i$-th pair is taken at time $t_i$ from recording $s_i$ and has embeddings $\mathbf z_i^{(\mathrm{Az})}$ and $\mathbf z_i^{(\mathrm{El})}$. The weight of candidate $j$ for anchor $i$ is
\begin{equation}
w_{ij}=\frac{\tilde w_{ij}}{\sum_{j'=1}^{B}\tilde w_{ij'}},
\qquad
\tilde w_{ij}=
\begin{cases}
\exp\!\big(-|t_i-t_j|/\sigma_w\big), & s_i=s_j \text{ and } |t_i-t_j|\le\Delta,\\
0, & \text{otherwise},
\end{cases}
\label{eq:soft_weights}
\end{equation}
where $t_i$, $\Delta$, and the decay rate $\sigma_w$ are measured in frames. The synchronized frame receives the largest weight. The temporal-weighted contrastive loss is
\begin{equation}
\mathcal{L}_{\mathrm{Az}\rightarrow\mathrm{El}} = -\frac{1}{B}\sum_{i=1}^{B}\sum_{j=1}^{B} w_{ij}\,
\log\frac{\exp\!\big(\mathbf{z}_i^{(\mathrm{Az})T}\mathbf{z}_j^{(\mathrm{El})}/\zeta\big)}
{\sum_{j'=1}^{B}\exp\!\big(\mathbf{z}_i^{(\mathrm{Az})T}\mathbf{z}_{j'}^{(\mathrm{El})}/\zeta\big)},
\label{eq:lcv_dir}
\end{equation}
where $\zeta$ is a temperature parameter~\cite{oord2018representation, chen2020simple}. The same loss is computed with the elevation view as the anchor, and the two directions are averaged:
\begin{equation}
\mathcal{L}_{\mathrm{cv}} = \tfrac{1}{2}\big(\mathcal{L}_{\mathrm{Az}\rightarrow\mathrm{El}}+\mathcal{L}_{\mathrm{El}\rightarrow\mathrm{Az}}\big).
\label{eq:lcv}
\end{equation}
This objective encourages cross-view consistency while preserving similarity between temporally neighboring observations.

\subsubsection{Temporal Masked Reconstruction}
\label{sec:temporal_masked_reconstruction}

Contrastive learning tells the backbone which observations should be similar, but not how a motion unfolds over time. Different activities can produce similar heatmaps at certain moments and differ mainly in how these heatmaps evolve. To capture this evolution, BeamLens hides part of a heatmap sequence and trains the model to recover it from the surrounding frames.

For each view, the shared backbone $f_e$ converts a sequence of $T$ consecutive heatmaps into features $\mathbf h_1^{(v)},\ldots,\mathbf h_T^{(v)}$. A continuous segment of these features, indexed by $\mathcal T_{\mathrm m}\subset\{1,\ldots,T\}$, is replaced with a learnable mask token~\cite{he2022masked}. A continuous segment, rather than isolated frames, is masked because adjacent frames are nearly identical and could otherwise simply be copied. After positional encodings are added, the sequence is fed into a temporal Transformer~\cite{vaswani2017attention}, in which each masked position aggregates information from all visible frames. A heatmap decoder then maps each masked position back to a heatmap $\widehat{\mathbf S}_t^{(v)}$, which is compared with the original using the mean squared error (MSE):
\begin{equation}
    \mathcal L_{\mathrm{rec}}
    = \frac{1}{2c_{\mathrm{rec}}}\sum_{v\in\{\mathrm{Az},\mathrm{El}\}}\operatorname{MSE}\big(\widehat{\mathbf S}_{\mathcal T_{\mathrm m}}^{(v)},\,\widetilde{\mathbf S}_{\mathcal T_{\mathrm m}}^{(v)}\big),
    \label{eq:temporal_reconstruction_loss}
\end{equation}
where the MSE is averaged over all masked positions and pixels. The constant $c_{\mathrm{rec}}$ is the reconstruction error obtained by copying the nearest visible frame into each masked position, estimated once from the training data. It normalizes the reconstruction loss relative to this simple temporal prediction baseline, helping balance its contribution to the overall training objective. The two views share the same Transformer and decoder but are reconstructed separately.

The two objectives are combined into the overall self-supervised learning (SSL) objective:
\begin{equation}
    \mathcal L_{\mathrm{SSL}} = \mathcal L_{\mathrm{cv}} + \beta\,\mathcal L_{\mathrm{rec}},
    \label{eq:ssl_total_loss}
\end{equation}
where $\beta$ balances the two objectives. $\mathcal L_{\mathrm{cv}}$ aligns the two views and keeps nearby moments close, while $\mathcal L_{\mathrm{rec}}$ requires the frame-level features to carry enough information to predict the motion over longer time spans.

After pretraining, only the shared backbone $f_e$ is kept; the projector, temporal Transformer, and heatmap decoder are discarded. The reconstruction task serves as an auxiliary objective that encourages $f_e$ to produce features carrying the information needed for temporal prediction, so downstream models do not need to rely on the pretrained Transformer. For each downstream task, $f_e$ encodes a sequence of consecutive heatmap pairs from the two views, and a task-specific head maps the resulting features to the task output, i.e., the activity class, the 3D human pose, or the human silhouette.
\section{Evaluation}
\label{sec:evaluation}

\subsection{Experimental Setup}
\label{sec:setup}

\subsubsection{Devices}
We implement BeamLens with two AP-STA links. Each link consists of a NETGEAR R7800 router as the AP and a laptop with an Intel AX210 NIC as the STA. Each AP has four transmitting antennas, and each STA has two receiving antennas. The four transmitting antennas of each AP are arranged at positions $\{0,1,4,6\}d$, and the antenna arrays of the two APs are oriented horizontally and vertically to provide the azimuth and elevation views, respectively. A third AX210-based laptop in monitor mode serves as a passive sniffer, which captures the BFI of both links using Wireshark. The reports of the two links are distinguished by their medium access control (MAC) addresses. Both links operate on channel 149 in the 5 GHz band with an 80 MHz bandwidth, and each BFI report covers the 234 data subcarriers of the channel.

To obtain the ground truth of the downstream tasks, an RGB camera records the participant at 30 frames per second, and we increase the channel sounding rate of each AP to 30 BFI reports per second to match the frame rate of the camera. The video is only used to generate the activity labels, 3D human poses by MotionBERT~\cite{zhu2023motionbert}, and human silhouettes by Self-Correction for Human Parsing (SCHP)~\cite{li2020self} as the ground truth. The BFI capture and the video recording are started simultaneously at the beginning of each recording session, and each heatmap is paired with the video frame closest to the center of its 1 s window.

\subsubsection{Environments}
\begin{figure*}[t]
\centering
\includegraphics[width=\textwidth]{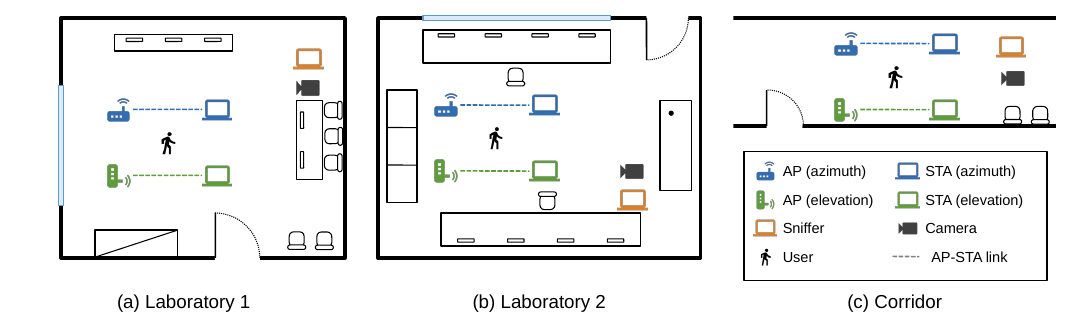}
\caption{Experimental environments and device deployments.}
\label{fig:environments}
\end{figure*}

We evaluate BeamLens in three indoor environments: two laboratories (6\,m $\times$ 5\,m and 6.3\,m $\times$ 5\,m) and a corridor (8\,m $\times$ 3\,m), as shown in Fig.~\ref{fig:environments}. In each environment, the two AP-STA links are placed in parallel, with a distance of 3\,m between the AP and the STA of each link and a separation of 2\,m between the two links. The participant performs activities in the area between the two links, at a distance of 1.5\,m from the line connecting the two APs, and the camera is placed to capture the participant in the LoS. Unless otherwise specified, these settings are used throughout the evaluation.

\subsubsection{Data Collection}
We recruit 8 participants (4 males and 4 females) of different heights, weights, and ages. Each participant performs 12 activities multiple times in each of the three environments. In addition, we collect unlabeled recordings for self-supervised pretraining, during which the participants move freely without following any activity protocol. The data collection spans four months, during which we collect over 1,000,000 BFI reports in total, yielding approximately 500,000 heatmap pairs. This study has been approved by the Institutional Review Board (IRB), and all participants are anonymized.

Among them, approximately 300,000 heatmap pairs come from the unlabeled free-movement recordings, and the remaining approximately 200,000 heatmap pairs come from the labeled activity recordings. The labeled recordings are randomly split into training, validation, and test sets with a ratio of 7:1:2, where each recording is assigned to only one set. The self-supervised pretraining uses the unlabeled heatmap pairs together with the training set of the labeled recordings, whose labels are not used. The validation and test sets are excluded from pretraining, so the test data are never seen by the backbone before evaluation. Unless otherwise specified, this split is used throughout the evaluation.

\subsubsection{Implementation Details}
\label{sec:implementation}
For each view, the covariance matrices of the BFI reports within a 1 s window are averaged and converted into a heatmap on a $201 \times 261$ AoD-RToF grid, covering AoD from $-90^\circ$ to $90^\circ$ and RToF from 0 to 10 ns. The window slides by one report, producing heatmap pairs at 30 Hz. The shared backbone $f_e$ is a lightweight four-stage convolutional neural network (CNN) that maps each heatmap to a 256-dimensional feature. During pretraining, each sequence contains $T=60$ consecutive frames (2 s), a continuous segment of 24 frames is masked, and we set $\Delta=15$, $\sigma_w=5$, and $\beta=0.5$. All models are trained with AdamW~\cite{loshchilov2017decoupled} and early stopping on the validation set. During fine-tuning, the backbone is updated with a learning rate of $3\times10^{-5}$.

\subsubsection{Downstream Tasks}
\label{sec:tasks}
For each downstream task, the features of the two views are concatenated per frame and fed into a task-specific head, which consists of a two-layer bidirectional gated recurrent unit (GRU)~\cite{cho2014learning} with a 60-frame input followed by a task-specific output layer.

\textit{Human activity recognition.} The participants perform 12 activities: bending, marching in place, jumping, torso rotation, arm swinging, lateral stepping, squatting, pushing and pulling, single-arm raising, forward lunging, side kicking, and walking. The head uses a GRU with a hidden size of 48 followed by a classifier, and is trained with the cross-entropy loss with a label smoothing of 0.1, a learning rate of $3\times10^{-4}$, a weight decay of $2\times10^{-2}$, and a batch size of 64 for up to 100 epochs.

\textit{3D human pose estimation.} We use MotionBERT~\cite{zhu2023motionbert} to estimate 17 pelvis-relative 3D body joints from the video. Since MotionBERT outputs poses in a normalized scale, we rescale them to metric units according to the measured body height of each participant, and use the rescaled poses as the ground truth. The head uses a GRU with a hidden size of 128 followed by a joint regressor that outputs the 3D coordinates of the 17 joints, and is trained with the MSE loss, a learning rate of $1\times10^{-3}$, a weight decay of $10^{-4}$, and a batch size of 32 for up to 50 epochs.

\textit{Human silhouette estimation.} We use SCHP~\cite{li2020self} to generate human segmentation masks from the video and convert them into binary silhouettes as the ground truth. The head uses a GRU with a hidden size of 128 followed by a mask decoder that outputs a $128\times128$ binary silhouette, and is trained with a combination of the cross-entropy loss and the Dice loss, a learning rate of $3\times10^{-4}$, a weight decay of $10^{-3}$, and a batch size of 32 for up to 30 epochs. Since the ground truth of 3D human pose estimation and human silhouette estimation is derived from vision models, the reported metrics measure the agreement with these vision-derived references.

\subsubsection{Baselines}

We compare BeamLens with four baselines. All of them use the same task-specific heads and the same data split as BeamLens, and differ in the input or in how the backbone is obtained.
\begin{itemize}
\item \textit{Constant Predictor}: It ignores the input and always outputs the same prediction derived from the training labels, i.e., the most frequent activity, the mean 3D pose, and the mean silhouette for the three tasks, respectively. It indicates the performance achievable without any sensing information.
\item \textit{Random Features}: It takes the AoD-RToF heatmaps as input, but its backbone is randomly initialized and kept frozen, and only the task-specific heads are trained. It indicates how much of the performance comes from the heads alone.
\item \textit{Raw-BFI}: It takes the beamforming matrices $\widetilde{\mathbf V}_k$ reconstructed from BFI, with their real and imaginary parts as separate channels, as input instead of the AoD-RToF heatmaps, and is pretrained and evaluated in the same way as BeamLens. It evaluates the benefit of the AoD-RToF heatmap representation.
\item \textit{Scratch}: It takes the same AoD-RToF heatmaps as BeamLens but trains the backbone and the heads together from scratch with labels only, without pretraining. It evaluates the benefit of self-supervised pretraining.
\end{itemize}
For BeamLens, we report two versions: \textit{Frozen}, where the pretrained backbone is fixed and only the heads are trained, and \textit{Fine-tuned}, where the backbone and the heads are trained together. Unless otherwise specified, BeamLens refers to the Frozen version in the following experiments.

\subsubsection{Evaluation Metrics}
For human activity recognition, we report the accuracy, i.e., the percentage of correctly classified samples, and the macro-F1 score, i.e., the F1 score computed for each activity class and then averaged over all classes. For 3D human pose estimation, we report the mean per-joint position error (MPJPE)~\cite{ionescu2013human3}, i.e., the average Euclidean distance between the predicted and ground-truth positions of the 17 joints after aligning their pelvis joints. We also report the Procrustes-aligned MPJPE (PA-MPJPE), which is computed after aligning the predicted pose to the ground-truth pose in translation, rotation, and scale, and thus focuses on the error in body posture. For human silhouette estimation, we report the IoU~\cite{everingham2010pascal} and the Dice~\cite{milletari2016v} score, both of which measure the overlap between the predicted silhouette $P$ and the ground-truth silhouette $G$:
\begin{equation}
\mathrm{IoU} = \frac{|P \cap G|}{|P \cup G|}, \qquad
\mathrm{Dice} = \frac{2|P \cap G|}{|P| + |G|},
\label{eq:iou_dice}
\end{equation}
where $|\cdot|$ denotes the number of pixels. IoU is the ratio of the overlapping area to the union of the two silhouettes, while Dice is the ratio of twice the overlapping area to the sum of their areas. Both metrics range from 0 to 1, and a higher value indicates a better match.

\subsection{Overall Performance}
\label{sec:overall}
We first evaluate the overall performance of BeamLens on the three downstream tasks and compare it with the four baselines. Table~\ref{tab:overall} summarizes the results.

\begin{figure}[t]
\centering
\includegraphics[width=0.55\textwidth]{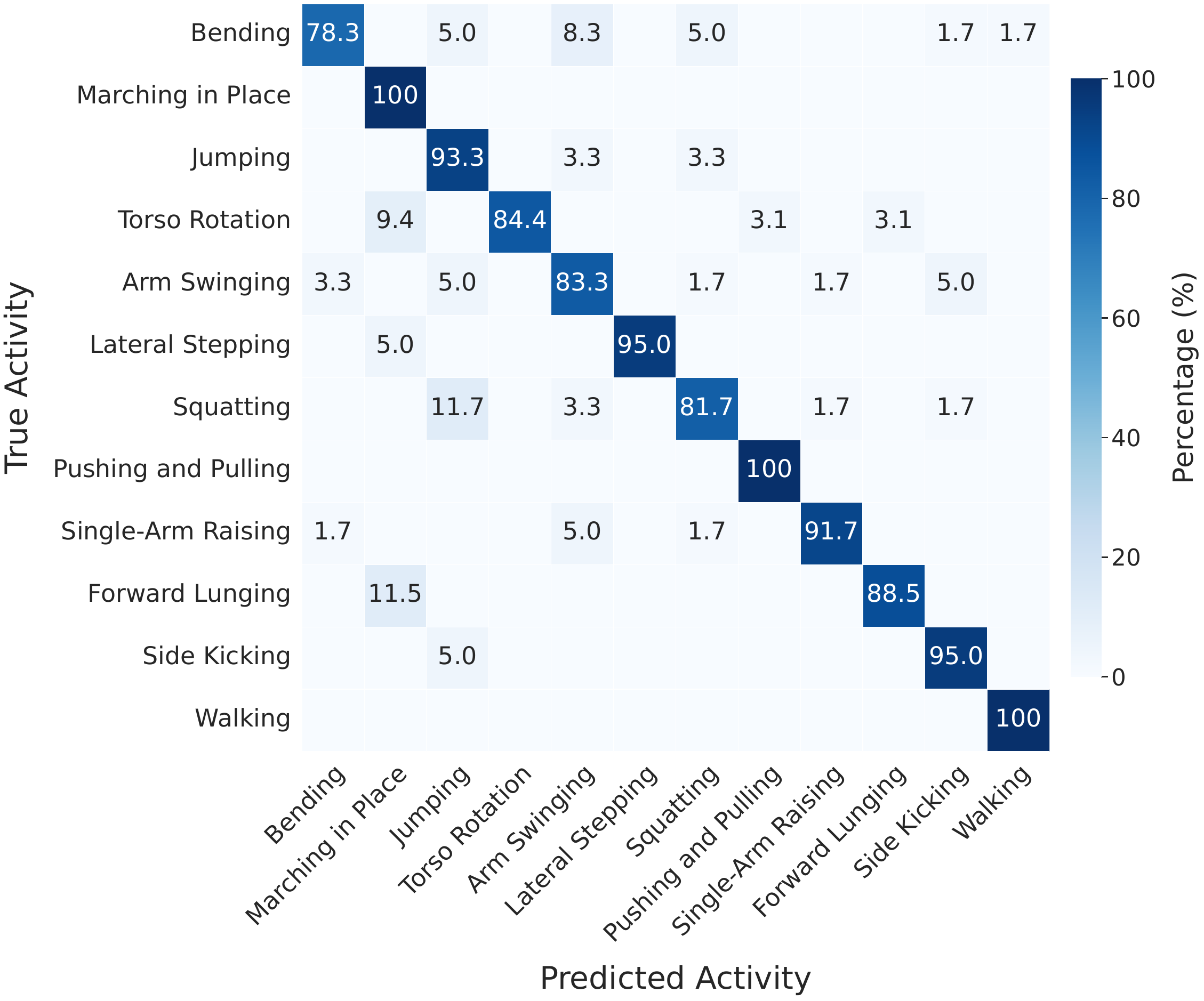}
\caption{Confusion matrix of human activity recognition with BeamLens (Frozen).}
\label{fig:confusion}
\end{figure}

\textit{Human activity recognition.} With the backbone frozen, BeamLens correctly recognizes 90.2\% of the activities, with a macro-F1 score of 91.0\%. Fine-tuning the backbone together with the task-specific head further improves the accuracy and the macro-F1 score to 91.1\% and 91.2\%, respectively. In comparison, the accuracies of Raw-BFI, Scratch, and Random Features are 87.1\%, 86.2\%, and 82.3\%, respectively, while Constant Predictor only reaches 10.7\%. As shown in Fig.~\ref{fig:confusion}, most activities are recognized with an accuracy above 85\%, and the main confusions occur between activities with similar body movements. For example, 11.7\% of squatting is recognized as jumping, since both involve bending the knees and moving the body up and down. Similarly, 11.5\% of forward lunging is recognized as marching in place, since both involve lifting and stepping the legs.

\begin{figure*}[t]
\centering
\includegraphics[width=\textwidth]{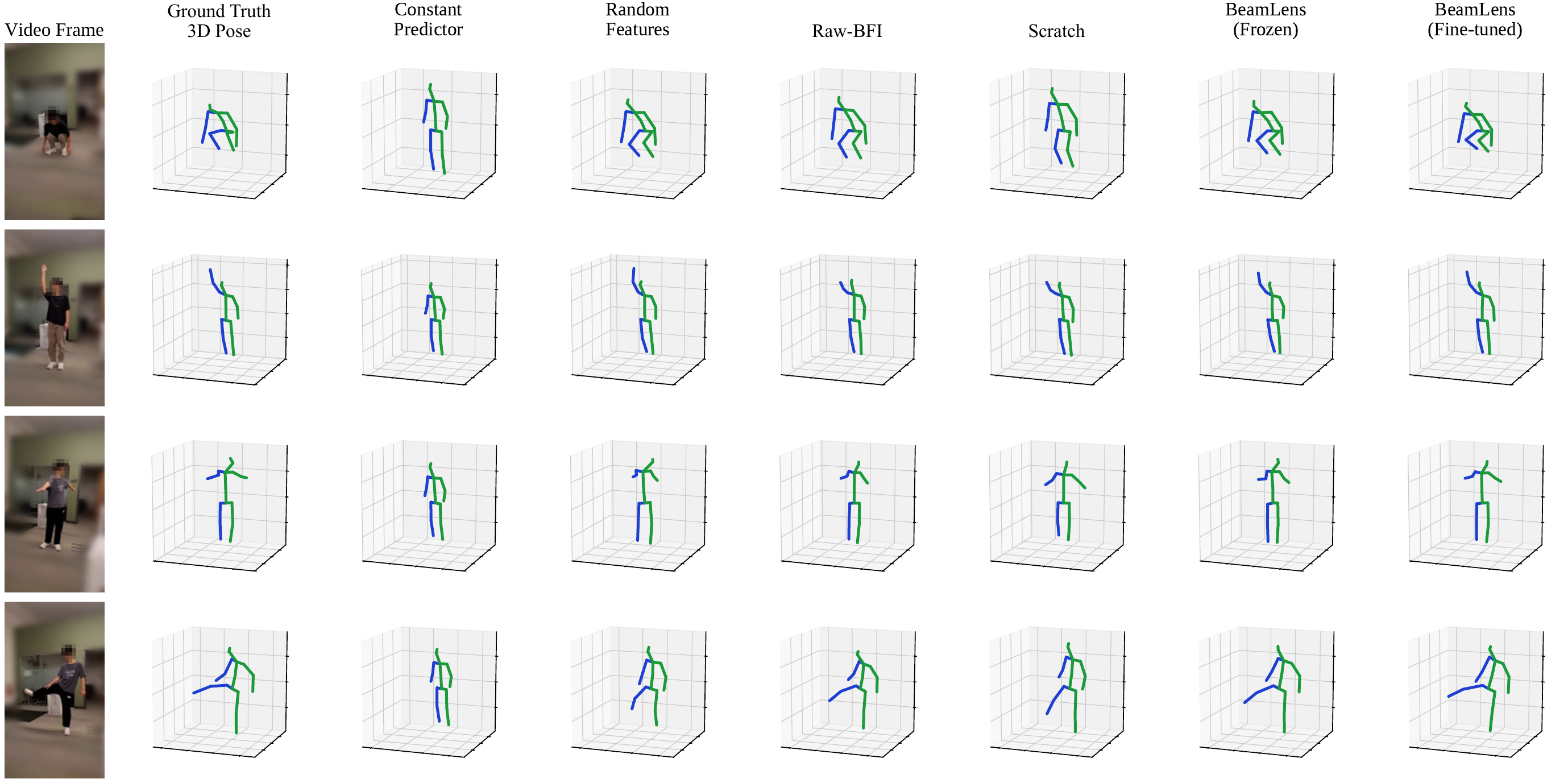}
\caption{Examples of 3D human pose estimation. Each row shows a video frame, the ground-truth 3D pose, and the 3D poses estimated by the baselines and BeamLens.}
\label{fig:pose_examples}
\end{figure*}

\textit{3D human pose estimation.} Among the baselines, Raw-BFI yields the lowest MPJPE of 9.1\,cm, followed by Scratch (9.4\,cm), Random Features (10.7\,cm), and Constant Predictor (12.9\,cm). BeamLens further reduces the MPJPE to 8.6\,cm with the backbone frozen and to 7.8\,cm with the backbone fine-tuned, and its PA-MPJPE drops to 5.4\,cm and 5.1\,cm, respectively. Fig.~\ref{fig:pose_examples} visualizes several estimated 3D human poses. The poses estimated by BeamLens are the closest to the ground truth. In particular, BeamLens accurately estimates the limbs, such as the raised arm and the kicking leg, even though their reflections are much weaker than those of the torso, whereas the baselines fail to recover these limb movements.

\begin{figure*}[t]
\centering
\includegraphics[width=\textwidth]{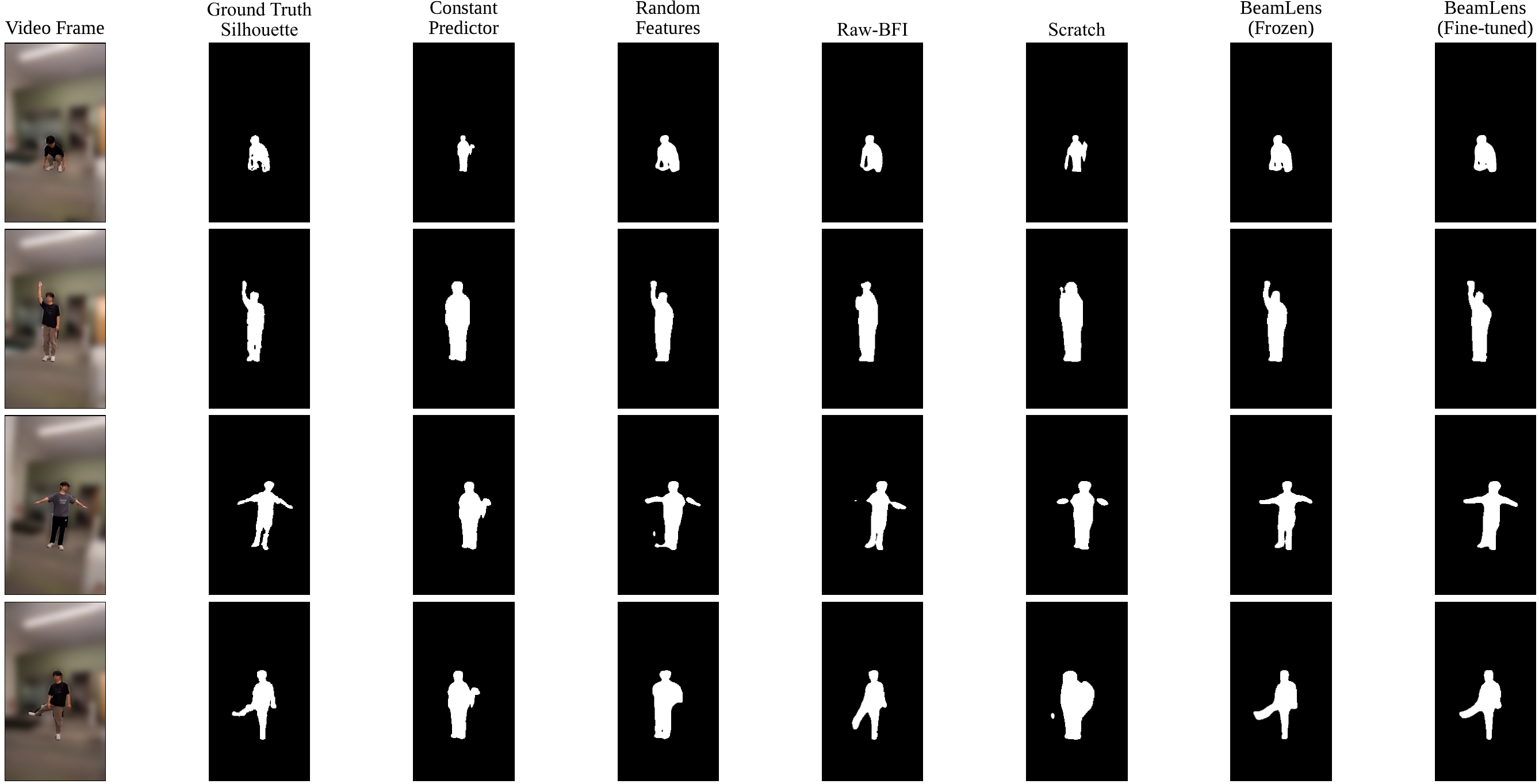}
\caption{Examples of human silhouette estimation. Each row shows a video frame, the ground-truth silhouette, and the silhouettes estimated by the baselines and BeamLens.}
\label{fig:silhouette_examples}
\end{figure*}

\textit{Human silhouette estimation.} Fig.~\ref{fig:silhouette_examples} compares the estimated silhouettes. BeamLens recovers the outstretched arms and the kicking leg, whereas the baselines tend to produce coarse, standing-like silhouettes. Quantitatively, the frozen BeamLens attains an IoU of 70.6\% and a Dice score of 82.8\%, and fine-tuning raises them to 72.1\% and 83.8\%. The IoUs of the baselines range from 65.6\% for Constant Predictor to 70.1\% for Raw-BFI.

Overall, BeamLens outperforms all baselines on all three tasks, even with the backbone frozen. This is because BeamLens extracts fine-grained spatial information about the human body from BFI, and fuses the two complementary views through self-supervised learning on unlabeled BFI, which yields a general-purpose backbone for versatile human sensing.

\begin{table*}[t]
\centering
\caption{Overall performance on the three downstream tasks.}
\label{tab:overall}
\small
\setlength{\tabcolsep}{4pt}
\begin{tabular}{l cc cc cc}
\toprule
& \multicolumn{2}{c}{Human activity recognition} & \multicolumn{2}{c}{3D human pose estimation} & \multicolumn{2}{c}{Human silhouette estimation} \\
\cmidrule(lr){2-3} \cmidrule(lr){4-5} \cmidrule(lr){6-7}
Method & Acc. (\%) $\uparrow$ & Macro-F1 (\%) $\uparrow$ & MPJPE (cm) $\downarrow$ & PA-MPJPE (cm) $\downarrow$ & IoU (\%) $\uparrow$ & Dice (\%) $\uparrow$ \\
\midrule
Constant Predictor & 10.73 & 1.62 & 12.91 & 7.26 & 65.61 & 79.24 \\
Random Features & 82.29 & 84.57 & 10.68 & 6.16 & 66.60 & 79.95 \\
Raw-BFI & 87.06 & 89.68 & 9.06 & 5.88 & 70.12 & 82.44 \\
Scratch & 86.23 & 89.07 & 9.39 & 5.72 & 69.64 & 82.10 \\
\midrule
BeamLens (Frozen) & 90.16 & 91.03 & 8.64 & 5.44 & 70.64 & 82.79 \\
BeamLens (Fine-tuned) & \textbf{91.06} & \textbf{91.18} & \textbf{7.84} & \textbf{5.06} & \textbf{72.07} & \textbf{83.77} \\
\bottomrule
\end{tabular}
\end{table*}

\subsection{Accuracy of AoD and RToF Estimation}
\label{sec:aod_accuracy}

\begin{table}[t]
\centering
\caption{Errors of AoD and RToF estimation over all positions and experiments.}
\label{tab:aod_accuracy}
\begin{tabular}{l c c c}
\toprule
 & Mean & Min & Max \\
\midrule
AoD error ($^\circ$) & 7.8 & 0.2 & 22.4 \\
RToF error (ns) & 1.2 & 0.1 & 5.3 \\
\bottomrule
\end{tabular}
\end{table}
The performance of BeamLens relies on whether the AoD-RToF heatmaps correctly locate the reflections from the human body. We therefore evaluate the accuracy of the AoD and RToF estimated from BFI.

To obtain the ground truth, the participant stands at 8 known positions in the sensing area, and we conduct five experiments at each position. At each position, we measure the distances between the participant and the AP and STA, from which the ground-truth RToF is computed. Since the human body is an extended reflector rather than a point, we define the ground-truth AoD by the boundaries of the body. In the azimuth view, the AoDs of the leftmost and rightmost parts of the body are computed from their measured positions relative to the AP using trigonometry. In the elevation view, the AoDs of the highest and lowest parts of the body, i.e., the head and the feet, are computed in the same way. Correspondingly, we take the AoD range of the human reflection in each background-removed heatmap, i.e., its minimum and maximum AoD, and compare its boundaries with the ground truth.

Table~\ref{tab:aod_accuracy} shows the results. Averaged over all positions and experiments, BeamLens achieves a mean AoD error of 7.8$^\circ$ and a mean RToF error of 1.2\,ns, with minimum errors of only 0.2$^\circ$ and 0.1\,ns. Larger errors mainly occur when the participant stands close to the AP, where the boundaries of the body approach $\pm 90^\circ$, since the angle estimation performance of a linear array degrades considerably near these angles~\cite{shirvani2011novel}. These results show that the heatmaps extracted from BFI reliably capture the spatial extent of the human body, providing the spatial information required by the downstream tasks.

\subsection{Impact of the Amount of Labeled Data}
\label{sec:label}

\begin{figure*}[t]
\centering
\includegraphics[width=\textwidth]{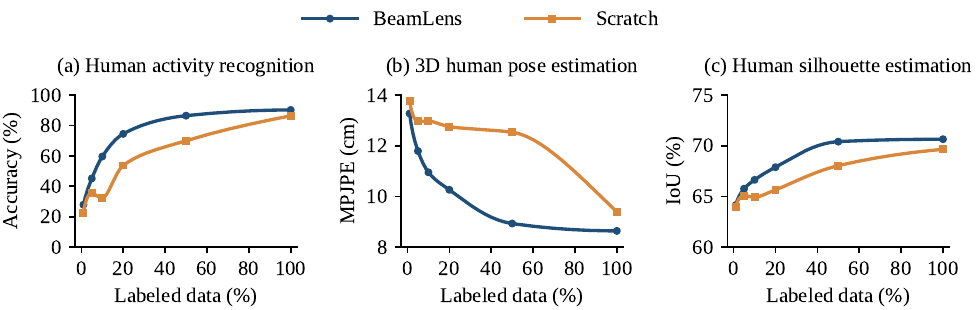}
\caption{Impact of the amount of labeled data.}
\label{fig:different_label}
\end{figure*}

Annotating Wi-Fi sensing data is costly, as it requires synchronized video and vision-based reference labels. If the pretrained backbone learns effective representations, only a small amount of labeled data should be needed to train the task-specific heads. To verify this, we use 1\%, 5\%, 10\%, 20\%, 50\%, and 100\% of the labeled training set to train the task-specific heads, while keeping the backbone of BeamLens frozen and the validation and test sets unchanged. Note that these ratios only refer to the labeled data used for training the task-specific heads. We compare BeamLens with Scratch, which trains its backbone together with the task-specific heads from scratch using the same labeled data.

Fig.~\ref{fig:different_label} plots the performance under different amounts of labeled data. The performance of BeamLens improves rapidly as the amount of labeled data increases, and saturates at around 50\%. For example, the accuracy of BeamLens increases from 27.9\% with 1\% of the labels to 86.4\% with 50\%, and only slightly further to 90.2\% with all labels. In contrast, the performance of Scratch improves gradually and keeps increasing until all labels are used. Notably, with only 50\% of the labels, BeamLens achieves an accuracy of 86.4\%, an MPJPE of 8.9\,cm, and an IoU of 70.4\%, which are comparable to or even better than those of Scratch trained with all labels, i.e., 86.2\%, 9.4\,cm, and 69.6\%, respectively. In other words, BeamLens requires only half of the labeled data to match the performance of Scratch.

This is because BeamLens learns motion representations from unlabeled BFI during pretraining, so the task-specific heads only need a small amount of labeled data to map these representations to the task outputs. In contrast, Scratch has to learn both the representations and the task mapping from the labeled data alone. These results demonstrate that BeamLens substantially reduces the labeling effort required for versatile human sensing.

\subsection{Impact of Different Environments}
\label{sec:eval_env}

\begin{figure*}[t]
  \centering
  \includegraphics[width=\textwidth]{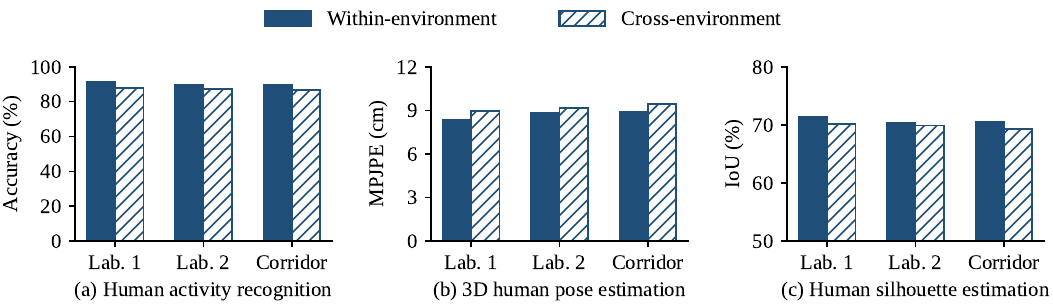}
  \caption{Impact of different environments. Within-environment: the model is trained on all three environments and tested on each environment separately. Cross-environment: the model is trained on two environments and tested on the held-out one.}
  \label{fig:impact_env}
\end{figure*}

Room geometry and furniture alter the multipath components captured in BFI, which may affect the performance of Wi-Fi sensing in different environments. We therefore evaluate the impact of different environments on BeamLens. We conduct the evaluation in the three environments in Fig.~\ref{fig:environments} under two settings. In the within-environment setting, we use the model in Section~\ref{sec:overall}, which is trained on all three environments, and report its test performance in each environment separately. In the cross-environment setting, we adopt leave-one-environment-out evaluation. For example, to test on Laboratory 1, the model is trained on Laboratory 2 and the corridor, without using any labeled data from Laboratory 1. The data of the held-out environment are also excluded from pretraining.

Fig.~\ref{fig:impact_env} presents the performance of BeamLens in each environment under the two settings. In the within-environment setting, BeamLens achieves accuracies of 91.3\%, 89.5\%, and 89.7\% in Laboratory 1, Laboratory 2, and the corridor, respectively. When tested in unseen environments, the accuracies are 87.8\%, 87.3\%, and 86.6\%, respectively. Similarly, the MPJPE increases by only 0.4-0.6\,cm, and the IoU decreases by only 0.4-1.2\%. The small performance gap between the two settings indicates that the environment has little impact on BeamLens. This is because BeamLens removes the environment-dependent reflections, such as those from the LoS path, walls, and furniture, through background subtraction, so the heatmaps mainly retain the reflections from the human body. In addition, self-supervised pretraining enables BeamLens to learn motion representations that are less sensitive to the environment. These results demonstrate that BeamLens generalizes well across different environments and can be deployed in new environments without additional labeling.

\subsection{Impact of Different Subjects}
\label{sec:eval_subject}

\begin{figure*}[t]
\centering
\includegraphics[width=\textwidth]{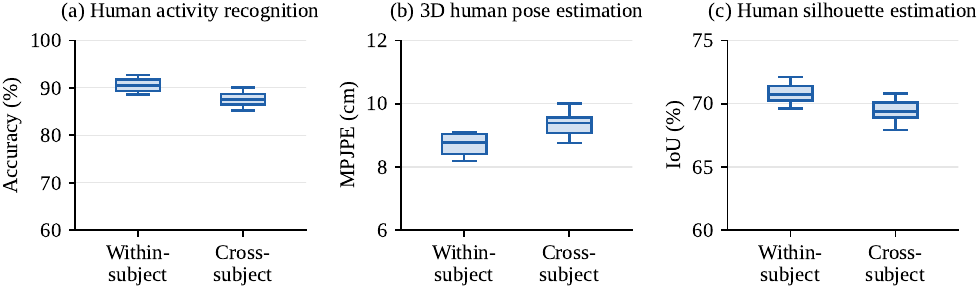}
\caption{Impact of different subjects. Within-subject: the model is trained with data from all subjects and tested on each subject separately. Cross-subject: the model is trained on seven subjects and tested on the held-out one.}
\label{fig:different_subject}
\end{figure*}

Different people have different body shapes and motion habits, which may affect the performance of Wi-Fi sensing. We therefore evaluate the impact of different subjects on BeamLens under two settings. In the within-subject setting, we use the default data split and report the test performance of each subject separately. In the cross-subject setting, we adopt leave-one-subject-out evaluation, where the model is trained on seven subjects and tested on the remaining one. The data of the held-out subject are also excluded from pretraining.

The performance of BeamLens for each subject is illustrated in Fig.~\ref{fig:different_subject}. In the within-subject setting, the accuracies of all subjects range from 88.6\% to 92.7\%. When tested on unseen subjects, the average accuracy decreases from 90.2\% to 87.2\%, the MPJPE increases by 0.6\,cm, and the IoU decreases by about 1\%. These small variations indicate that different subjects have little impact on BeamLens. This is because BeamLens learns motion representations from the unlabeled BFI of different subjects during pretraining, which capture how the human body moves rather than the body shape or motion habits of a specific person. These results demonstrate that BeamLens generalizes well to new users without collecting their labeled data.

\subsection{Impact of Different Distances}
\label{sec:eval_distance}

\begin{figure*}[t]
\centering
\includegraphics[width=\textwidth]{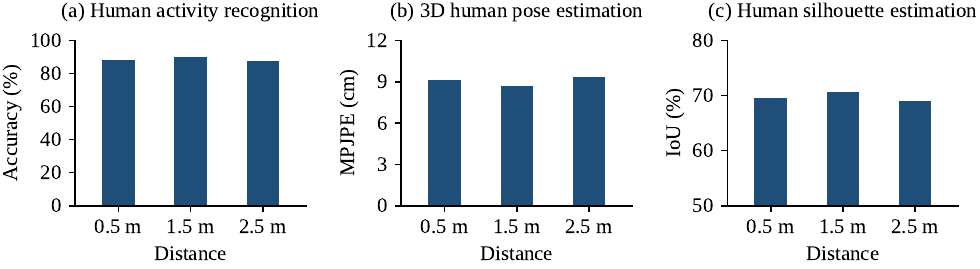}
\caption{Impact of the distance between the participant and the line connecting the two APs.}
\label{fig:different_distance}
\end{figure*}

The position of the user relative to the Wi-Fi devices varies in practice. We therefore evaluate the impact of the distance between the participant and the line connecting the two APs. In addition to the default distance of 1.5\,m, we collect data with the participant at 0.5\,m and 2.5\,m, while keeping the other settings unchanged.

We report the performance of BeamLens at the three distances in Fig.~\ref{fig:different_distance}. Compared with the default distance of 1.5\,m, the accuracy decreases from 90.2\% to 88.4\% and 87.8\% at 0.5\,m and 2.5\,m, respectively. Similarly, the MPJPE increases from 8.6\,cm to 9.1\,cm and 9.3\,cm, and the IoU decreases by about 1.5\%. These slight changes demonstrate that BeamLens maintains stable performance at different distances and is robust to the user position.

\subsection{Impact of LoS and NLoS Conditions}
\label{sec:eval_nlos}

\begin{figure*}[t]
\centering
\includegraphics[width=\textwidth]{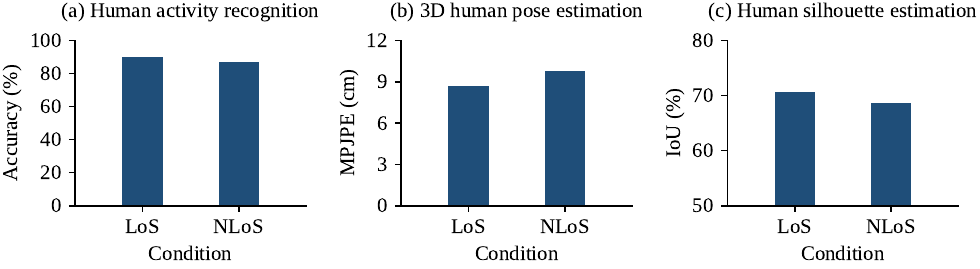}
\caption{Impact of LoS and NLoS conditions. In the NLoS condition, a wooden board blocks the direct path between the AP and the STA of each link.}
\label{fig:los_nlos}
\end{figure*}

In practical indoor deployments, the direct path between the AP and the STA may be blocked by partitions or furniture. We therefore evaluate BeamLens under both LoS and non-line-of-sight (NLoS) conditions. In the NLoS condition, we place a wooden board between the AP and the STA of each link to block the direct path, while keeping the other settings unchanged.

Fig.~\ref{fig:los_nlos} compares the performance of BeamLens under the LoS and NLoS conditions. When the direct path is blocked, the accuracy decreases from 90.2\% to 86.8\%, the MPJPE increases from 8.6\,cm to 9.7\,cm, and the IoU decreases by about 2\%. The performance degradation is small, indicating that BeamLens remains effective even when the direct path between the AP and the STA is blocked.

\subsection{Ablation Study}
\label{sec:eval_ablation}

\begin{table*}[t]
\centering
\caption{Ablation study. Each configuration removes or replaces one component of BeamLens while keeping the others unchanged. All configurations use the frozen backbone.}
\label{tab:ablation}
\small
\setlength{\tabcolsep}{4pt}
\begin{tabular}{l cc cc cc}
\toprule
 & \multicolumn{2}{c}{Human activity recognition} & \multicolumn{2}{c}{3D human pose estimation} & \multicolumn{2}{c}{Human silhouette estimation} \\
\cmidrule(lr){2-3} \cmidrule(lr){4-5} \cmidrule(lr){6-7}
Configuration & Acc. (\%) $\uparrow$ & Macro-F1 (\%) $\uparrow$ & MPJPE (cm) $\downarrow$ & PA-MPJPE (cm) $\downarrow$ & IoU (\%) $\uparrow$ & Dice (\%) $\uparrow$ \\
\midrule
Azimuth view only & 81.40 & 83.08 & 9.63 & 6.02 & 68.35 & 81.20 \\
Elevation view only & 71.91 & 74.45 & 10.42 & 6.68 & 66.82 & 80.11 \\
Contrastive only & 89.45 & 90.56 & 8.79 & 5.57 & 70.13 & 82.44 \\
Reconstruction only & 88.91 & \textbf{91.24} & 8.91 & 5.63 & 69.87 & 82.26 \\
ULA layout & 86.59 & 87.82 & 8.95 & 5.81 & 68.76 & 81.49 \\
\midrule
BeamLens (Frozen) & \textbf{90.16} & 91.03 & \textbf{8.64} & \textbf{5.44} & \textbf{70.64} & \textbf{82.79} \\
\bottomrule
\end{tabular}
\end{table*}

To evaluate the contribution of each component of BeamLens, we conduct an ablation study by removing one component at a time while keeping the others unchanged. Specifically, we consider the following configurations. (1) \textit{Azimuth view only} and \textit{Elevation view only} use the heatmaps of a single view. Since the cross-view contrastive learning requires two views, only the temporal masked reconstruction is used for pretraining. (2) \textit{Contrastive only} and \textit{Reconstruction only} pretrain the backbone with only the cross-view contrastive loss or only the temporal masked reconstruction loss, respectively. (3) \textit{ULA layout} arranges the transmitting antennas at $\{0,1,2,3\}d$ instead of the sparse physical antenna array at $\{0,1,4,6\}d$, and the uniform virtual antenna array is therefore not constructed. All configurations use the frozen backbone and the same data split. The results are shown in Table~\ref{tab:ablation}.

\textit{Two-view fusion.} Using a single view leads to the largest degradation. The accuracy decreases to 81.4\% with only the azimuth view and to 71.9\% with only the elevation view. This is because most of the activities mainly involve horizontal movements captured by the azimuth view, whereas the elevation view mainly captures vertical movements, so neither view alone fully describes the human motion.

\textit{Pretraining objectives.} Using only the contrastive loss or only the reconstruction loss reduces the accuracy to 89.5\% and 88.9\%, respectively, and increases the MPJPE to 8.8\,cm and 8.9\,cm. Although Reconstruction only achieves a slightly higher macro-F1 score (91.2\% vs. 91.0\%), the full model performs better on all other metrics, indicating that the two objectives are complementary: the contrastive loss aligns the two views at each moment, while the reconstruction loss captures how the motion evolves over time.

\textit{Sparse physical and uniform virtual antenna arrays.} Replacing the sparse physical antenna array with the ULA reduces the accuracy from 90.2\% to 86.6\% and increases the MPJPE from 8.6\,cm to 9.0\,cm. This is because the sparse physical antenna array enlarges the aperture from $3d$ to $6d$, and the uniform virtual antenna array enlarges the spatial dimension, which together improve the angular resolution of the heatmaps.

Overall, removing any component degrades the performance of BeamLens, demonstrating that the two-view fusion, the two pretraining objectives, and the antenna array design all contribute to versatile human sensing.
\section{Conclusion}
\label{sec:conclusion}

This paper presents BeamLens, a BFI-based Wi-Fi sensing system that extracts fine-grained spatial information to enable versatile human sensing. BeamLens jointly estimates AoD and RToF from BFI to extract fine-grained spatial information and leverages two AP-STA links to capture complementary azimuth and elevation views. To enhance spatial resolution, we design a sparse physical antenna array and construct a uniform virtual antenna array. Furthermore, we develop a self-supervised representation learning framework that integrates complementary spatial information across views and captures temporal dependencies to learn a general-purpose backbone for diverse human sensing tasks. Extensive experiments demonstrated the effectiveness and versatility of BeamLens, achieving 90.16\% accuracy in human activity recognition, an MPJPE of 8.64\,cm in 3D human pose estimation, and 70.64\% IoU in human silhouette estimation. BeamLens also demonstrated strong generalization, highlighting its potential for practical and versatile Wi-Fi-based human sensing.

%In this paper, we presented BeamLens, a system that extracts fine-grained spatial information from BFI for multi-task Wi-Fi sensing. BeamLens jointly estimates AoD and RToF from BFI by exploiting the spatial diversity across transmitting antennas and the frequency diversity across subcarriers, and captures complementary azimuth and elevation views with two AP-STA links. To overcome the limited number of antennas on commodity devices, BeamLens rearranges the existing antennas into a sparse array to enlarge the aperture and constructs a virtual array to enlarge the subspace available for path separation. BeamLens further fuses the two views through self-supervised learning with temporal-aware cross-view contrastive learning and temporal masked reconstruction, yielding a general-purpose backbone for multiple downstream tasks. We implemented BeamLens on commodity Wi-Fi devices and evaluated it with eight participants in three environments. With the pretrained backbone frozen, BeamLens achieves an activity recognition accuracy of 90.16\%, an MPJPE of 8.64\,cm for 3D human pose estimation, and an IoU of 70.64\% for human silhouette estimation, and it maintains its advantage over the baselines with fewer labels, in unseen environments, and on unseen subjects.

\bibliographystyle{plain}
\bibliography{ref}

\end{document}